\documentclass[%
reprint,
superscriptaddress,
 amsmath,amssymb,
 aps,
 pre
]{revtex4-1}
\usepackage[english]{babel}
\usepackage{graphicx}
\usepackage{dcolumn}
\usepackage{bm}

\usepackage{amsmath, amsthm, amssymb, amsfonts}
\usepackage{lmodern} 
\usepackage[utf8]{inputenc}
\usepackage{mathtools}
\usepackage{hyperref}
\usepackage{graphicx}
\usepackage{dsfont}
\usepackage{fullpage}
\usepackage{color}
\usepackage{soul} 
\usepackage[title]{appendix}
\usepackage{tikz}
\usepackage{stmaryrd}
\usetikzlibrary{patterns}
\usetikzlibrary{shapes.multipart}
\usetikzlibrary{arrows}

\graphicspath{{data/}}

\definecolor{labelkey}{cmyk}{.4,.2,0,0}

\newcommand{\be}{\begin{equation}}
\newcommand{\ee}{\end{equation}}
\newcommand{\bea}{\begin{eqnarray}}
\newcommand{\eea}{\end{eqnarray}}

\newcommand{\Tr}{\mathrm{Tr}}

\newcommand{\Var}{\mathrm{Var}}

\newcommand{\R}{\ensuremath{\mathbb{R}}}

\newcommand{\Z}{\ensuremath{\mathbb{Z}}}

\newcommand{\I}{\ensuremath{\mathbf{i}}}

\renewcommand{\rho}{\varrho}

\newcommand{\eps}{\varepsilon}
\renewcommand{\leq}{\leqslant}
\renewcommand{\geq}{\geqslant}

\renewcommand{\ge}{\geqslant}

\theoremstyle{definition}
\newtheorem{remark}{Remark}[section]

\begin{document}



\title{Random walks in Dirichlet random environment in dimension $d+1$}

\author{Guillaume Barraquand}
\affiliation{Laboratoire de Physique de l'\'Ecole Normale Sup\'erieure, École normale supérieure, Université PSL, CNRS, Sorbonne Université, Université Paris-Cité, 24 rue Lhomond, 75005 Paris, France} 

\author{Alexander K. \surname{Hartmann}}
\affiliation{Institut f{\"u}r Physik, Universit{\"a}t Oldenburg, 26111 Oldenburg, Germany}

\author{Pierre Le Doussal}
\affiliation{Laboratoire de Physique de l'\'Ecole Normale Sup\'erieure, École normale supérieure, Université PSL, CNRS, Sorbonne Université, Université Paris-Cité, 24 rue Lhomond, 75005 Paris, France}

\date{\today}

\begin{abstract}
The atypical behaviour of random walks in time-dependent random environment was recently related to Kardar-Parisi-Zhang (KPZ) growth. While this is now well-understood in spatial dimension $d=1$, further efforts are  necessary to better understand these connections in dimensions $d>1$. In this paper, we study this problem numerically for $d=1, 2$ and $3$, focusing on a discrete model with Dirichlet distributed transition probabilities. This model is a  generalization of an integrable model in $d=1$, and it  has the advantage of admitting an explicit, product-form, stationary measure. We verify that the growth of the variance of the logarithm  of point-to-point probabilities, namely from the origin to position $x$ in time $t$, is compatible with KPZ growth in dimension $d=1$ and $d=2$. In spatial dimension $d=3$, we confirm the existence of a phase transition as the angle $\vert x\vert /t$ increases and we obtain a lower bound based on an exact second moment calculation. We find that in the weak disorder phase the 
point-to-point probability acquires a heavy tailed distribution, and 
that in the strong disorder phase the cumulants of its logarithm grow with time. 
Further, we show that for this model, we can compute exactly the sample to sample variance of the thermal average $\overline{ \langle x \rangle^2}$ and that it is related to the extreme diffusion coefficient introduced recently.
\end{abstract}

\maketitle


\section{Introduction} \label{sec:intro} 

\subsection{Overview}
Random walks in short-range correlated time-dependent random environments (RWRE)
have received recent renewed attention 
due to a relation to Kardar-Parisi-Zhang (KPZ) growth. 
Trajectories, starting from the origin and consisting of $t$ steps, are typically described by standard Gaussian diffusion,
with wandering $|x| \sim \sqrt{t}$. Atypical trajectories which reach points further away, i.e
with $|x| \gg \sqrt{t}$,
can be
described in analogy with directed polymers (DP) in random environments, a problem
related to KPZ growth \cite{kardar1986}. A solvable model in space dimension $d=1$, called 
Beta RWRE, was introduced in \cite{barraquand2017random}. For this model it was proved 
that the logarithm of the probability of reaching a point $x \sim v t$ ( with $v$ = const.) fluctuates from environment to environment
with the Tracy-Widom (TW) distribution.
The TW distribution, introduced to
describe extreme eigenvalues of random matrices \cite{tracy1994level},
is characteristic of
the large-scale behavior of KPZ growth.
It was then argued \cite{ledoussal2017diffusion} that the connection
between diffusion in time-dependent random media and growth 
extends to the KPZ equation itself, i.e., at finite time,
and in any space dimension $d$. In $d=1$ these finite time KPZ fluctuations
can be observed in the intermediate regime $|x| \sim t^{3/4}$
\cite{ledoussal2017diffusion,thiery2016exact,barraquand2020moderate}.
A number of recent mathematical works in space dimension $d=1$ 
\cite{corwin2017kardar,barraquand2020large,das2024multiplicative,das2024kpz, parekh2024hierarchy,   barraquand2025convergence}
have established a detailed description of the connection to the KPZ equation. 
An interesting corollary of these results is that 
the typical fluctuations of the positions of outliers in a cloud of $N \gg 1$ independent walkers,
e.g., in $d=1$ of the rightmost particle, 
in the same environment are, under proper scaling of $N$ and $t$,
related to the fluctuations described by the KPZ equation \cite{barraquand2017random, ledoussal2017diffusion, barraquand2020moderate, hass2023anomalous, hass2024extreme, hass2024first, hass2025super, hass2025universal}.

At present there are very few studies of this connection in $d>1$. The main prediction of \cite{ledoussal2017diffusion}
is that the KPZ growth exponent should be observed in atypical trajectories of RWRE's in any $d$,
and that a phase transition should occur in $d=3$. 
Indeed, since the KPZ growth/DP problem exhibits a transition from a weak noise phase
to a strong noise phase in $d=3$  \cite{kardar1986, monthus2006freezing} (see also  \cite{junk2024strong, junk2025coincidence} for the mathematical state of the art), an analogous transition should be 
observed in RWRE's in $d=3$ when $v$ increases beyond some 
threshold. Furthermore some predictions for the location of the  intermediate deviation
regime were obtained in $d=2$. Some of these predictions have been confirmed recently in \cite{drillick2025random} where, in the weak disorder phase, the fluctuations of the quenched random walk probability distribution are fully characterized in  terms of stochastic linear partial differential equations. In $d=2$, \cite{ark2025universal} studied numerically the universal scaling behaviour of the  probability that the random walker escapes a large disk.

The aim of this paper is to numerically test the connections between RWRE and KPZ on a discrete model of random walks in Dirichlet distributed random environment which naturally generalizes the Beta RWRE model from \cite{barraquand2017random}. We also show that the probability of reaching a point $\vert x\vert \sim \sqrt{t}$ can be very precisely analyzed, thanks to an explicit stationary measure, which facilitates the comparison between theoretical predictions and  numerical data. 

\subsection{Model} 

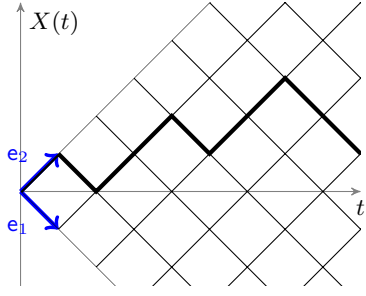
\begin{figure}
    \centering
    \begin{tikzpicture}[scale=0.5]
    \draw[-stealth', gray] (0,-2.5) -- (0,5) node[black, below right]{$X(t)$};
    \draw[-stealth', gray] (0,0) -- (9,0) node[below, black]{$t$};
    \clip (-1,-2.5) rectangle (9,5);
    \draw[->, blue, ultra thick] (0,0) -- (1,1) node[midway, above left]{$\mathsf{e}_2$};
   \draw[->, blue, ultra thick] (0,0) -- (1,-1) node[midway, below left]{$\mathsf{e}_1$};
\begin{scope}[rotate=-45, scale=sqrt(2)]
\clip (0,0) rectangle (7.5,7.5);
        \draw (0,0) grid (8,8);
        \draw[ultra thick] (0,0) -- (0,1) -- (1,1) -- (1,3) -- (2,3) -- (2,5) -- (4,5);
\end{scope}
    \end{tikzpicture}
    \caption{Space-time trajectory of a RWRE in dimension $1+1$, embedded in $\mathbb Z^2$.}
    \label{fig:d=1}
\end{figure}

Let us first consider a random walk in a random environment 
in dimension $d=1$ which makes jumps $\pm 1$.
We consider a random walker at position $X(t)$, where $t$ is the time. If for some time $t$, $X(t)=x$, we assume that $X(t+1)=x+1$ with probability $p_{x,t}$ and $X(t+1)=x-1$ with probability $1-p_{x,t}$. The $p_{x,t}$ are random variables that constitute the environment. The space-time trajectory, that is the graph of $t\mapsto X(t)$, may be seen as a walk in $\Z^2$ which makes steps along the basis vectors $\mathsf e_1, \mathsf e_2$ (see Figure \ref{fig:d=1}). In higher dimensions, we will likewise assume that the space-time trajectory makes steps along the basis vectors of $\Z^{d+1}$. 

 More precisely, we consider a random walk on $\mathbb Z^{d+1}$ starting from the origin. 
 We denote the possible space-time positions as $\mathsf n= (n_1,\dots,n_{d+1})$. If a walker is at $\mathsf n$ it randomly jumps to $\mathsf n+\mathsf e_i$, where $\mathsf e_i$, $1\leq i\leq d+1$, is one of the basis vectors. Since the walker starts at the orginin, $t=n_1+\dots+n_{d+1}$ is the number of steps. 
 The probability to jump to $\mathsf n+\mathsf e_i$ we denote as    $p_{\mathsf n,t}^i$ (the subscript $t$ in the notation is redundant but more along standard notation). We assume that the probability vectors 
$\mathsf p_{\mathsf n,t}=(p_{\mathsf n, t}^1,\ldots p_{\mathsf n,t}^{d+1})$, with $\sum_{i=1}^{d+1} p_{\mathsf n,t}^i=1$,
are independent and identically distributed as $t$ and $\mathsf n$ varies. The collection of random variables  $(\mathsf p_{\mathsf n})_{\mathsf n\in \Z_+^{d+1}}$ is called the environment. 
Since the environment is indexed by space-time points, the random walk sees independent jump probabilities at each time step. 
 We denote by $\mathsf R(t)\in \Z^{d+1}$ the space-time position of the walker, starting from $\mathsf R(0)=0$.   As in Figure \ref{fig:d=1}, we interpret the direction $(1,\dots,1)$ as the time axis. 
 
 We will be interested in the spatial position of the walker, i.e.,the  projection of $\mathsf R(t)$ on the hyperplane $\mathbb T_d$ orthogonal to the time axis , 
\be
\mathsf X(t) = \mathsf R(t)-t\mathsf d
\ee
where $\mathsf d=\frac{1}{d+1}(1,\dots, 1)$. If we let $\mathsf s_i=\mathsf e_i-\mathsf d$, the random walk $\mathsf X(t)$ makes steps in $d+1$ directions $\mathsf s_1, \dots, \mathsf s_{d+1}$,
where $\sum_i \mathsf s_i=0$. Another way to relate the space-time position $\mathsf R$ and the spatial position $\mathsf X$ is that if $\mathsf R(t) = (n_1, \dots, n_{d+1})$ then 
\be \label{defni}
\mathsf X(t) = \sum_{i=1}^{d+1} n_i\mathsf s_i.
\ee 
The probability distribution of the walker will be denoted by 
\be \label{defPn}
P(\mathsf n) = \mathbf P( \mathsf R(t)=\mathsf n)
\ee
where $\mathbf P$ denotes the probability on an endpoint
obtained by the measure on random walk paths, in some given environment. 

In the following we will distinguish the expectation values with respect to the random walk
in a given environment, denoted $\langle \cdots \rangle$, and the expectation values
with respect to the environment denoted $\overline{\rule{0pt}{2.5mm}\cdots}$.

\subsection{Known results} 

\paragraph*{Annealed random walk.} The annealed random walk is defined as the walk with transition probabilities $\overline{p^i}$. Let
\be 
\mathsf v_{\circ} = \overline{\langle \mathsf X(1)\rangle} = \sum_{i=1}^{d+1} \mathsf s_i \overline{p^i}
\ee 
 be the annealed drift. For all $t$, one has $ \overline{\langle X(t)\rangle} = \mathsf v_{\circ} t$. The diffusion matrix of the annealed random walk is such that for all $t$, 
\be 
\overline{\langle (\mathsf X(t)-\mathsf v_{\circ} t) (\mathsf X(t)-\mathsf v_{\circ} t)^T\rangle} = \mathsf D \, t.
\ee 
where $\mathsf D$ is the $(d+1)\times (d+1)$ matrix  
\be 
\mathsf D= \sum_{i=1}^{d+1} \mathsf s_i \mathsf s_i^T \overline{p^i}- \mathsf  v_0 \mathsf  v_0^T.
\ee
which in coordinates reads $\mathsf D_{a,b} =  \mathds{1}_{a=b} \overline{p^a} - \overline{p^a} \; \overline{p^b}$ and is of rank $d$ since $\mathsf X(t)$ belongs to
the hyperplane $\mathbb{T}_d$. One has also 
\be 
\overline{\langle (\mathsf X(t)-\mathsf v_{\circ} t)^2 \rangle } 
= D \, t ~ , ~  D= \Tr \, \mathsf D = \sum_{i=1}^{d+1} \overline{p^i} (1- \overline{p^i}) 
\ee

\paragraph*{Quenched random walk.}
For almost every sample of the environment, the scaled random walk $\eps (\mathsf X(t\eps^{-2})- \mathsf v_{\circ}t\eps^{-2}  )$ converges as $\eps\to 0$ to a $d$-dimensional Brownian motion  $\mathsf B(t)$ with the same diffusion matrix $\mathsf D$ \cite{rassoul2005almost, boldrighini1997almost}.
In particular, $\frac{1}{\sqrt{t}}\mathsf X(t)$ converges to a Gaussian 
distributed random variable. 

Under mild hypotheses on the environment distribution, random walk paths further satisfy a quenched 
large-deviation principle \cite{rassoul2013quenched}
\be 
\frac{1}{t}\log \mathbf P(\mathsf X(t)=t \mathsf v) \xrightarrow[t\to\infty]{} - I(\mathsf v), 
\label{eq:quenchedLDP}
\ee 
which gives the probability of very atypical random walks, and
where the rate function $I(\mathsf v)$ is the convex conjugate of the cumulant generating function, i.e.,
\be 
I(\mathsf v) = \sup_{\zeta\in \R^d} \lbrace \zeta\cdot \mathsf v -\lambda(\zeta)\rbrace
\ee 
where the function $\lambda$ is defined as the almost sure limit 
\be 
\lambda(\zeta) := \lim_{t\to\infty} \frac{1}{t} \log \langle e^{\zeta \mathsf X(t)} \rangle.
\ee 

The LDP \eqref{eq:quenchedLDP} is quenched in the sense that the left hand side is random and has subleading fluctuations depending on the environment. A consequence is that if one considers $N=e^{t I(\mathsf x)}$ independent random walks sampled in the same environment, there is a non trivial probability that at least one of these random walks is near $t\mathsf x$. To say something more precise, one needs to understand the fluctuations of $\log P(\mathsf X(t)=t \mathsf x)$ around its limit. 

In space dimension $d=1$, and if the random variables $\mathsf p_{\mathsf n}$ are distributed as independent Beta variables, the subleading fluctuations are such that
\be \label{TW}
\frac{\log \mathbf P(\mathsf X(t)= v t) + I(v)t}{\sigma(v)t^{1/3} } \xrightarrow[t\to\infty]{} \chi_2
\ee 
 where $\chi_2$ is a Tracy-Widom (TW) distributed random variable \cite{barraquand2017random, thiery2016exact, korotkikh2022hidden} and $\sigma(v)$ is an explicit function of $v$. The connection to Kardar-Parisi-Zhang universality class becomes even more apparent when one considers moderate deviations where $v$ is scaled as $t^{-1/4}$, i.e.,$X(t)\sim t^{3/4}$. More precisely, if $\mathsf p\sim\mathrm{Beta}(\alpha, \alpha)$, 
\begin{multline} 
\log \mathbf P\left(\frac{\mathsf X(t)}{\sqrt D}\geq \tau^{1/4} t^{3/4}\right) + \frac{\sqrt{\tau t}}{2} + \frac 1 4 \log \frac{\alpha^4t}{\tau} + \frac{\tau}{12} \\  \xRightarrow[t\to\infty]{} h(0,\tau/(2\alpha^2)) \label{moderatedev}
\end{multline} 
where $h(x,t)$ solves the KPZ equation with droplet initial condition \cite{ledoussal2017diffusion, barraquand2020moderate, das2024multiplicative}. 
As $\tau$ increases the random variable in the r.h.s. interpolates in distribution 
between the Gaussian Edwards Wilkinson statistics and the above TW distribution \eqref{TW}.

In the limit of small $\tau$ there is a crossover
from moderate deviations to typical fluctuations. The large-time limit of cumulants 
of $\log \mathbf P\left(\frac{\mathsf X(t)}{\sqrt D}\geq y  t^{1/2}\right)$
have been 
computed in \cite{krajenbrink2023crossover,hartmann2024probing}. In particular, it implies that 
\begin{equation} 
\frac{ \log \mathbf P\left(\frac{\mathsf X(t)}{\sqrt D}\geq y  t^{1/2}\right) - \int_{y}^{\infty}\frac{e^{-x^2/2}}{\sqrt{2\pi}} dx}{ \sqrt{c_2(y) t^{-1/2} }} \xRightarrow[t\to\infty]{} \mathcal N 
\end{equation}
where $\mathcal N$ is a standard Gaussian variable and $c_2(y)$ is given in 
(38) in \cite{hartmann2024probing} with 
$c_2(y)\sim y^2$ as $y\to\infty$, which matches with the small $\tau$ limit of the variance in \eqref{moderatedev}. 
As a space-time process, this Gaussian variable is the solution of an additive noise stochastic heat equation given in \cite[Theorem 1.13, bulk]{drillick2025random}.

\paragraph*{Higher dimensions.}

In dimensions $d=2$ and higher it was argued within a continuum model in \cite{ledoussal2017diffusion}
that the fluctuations of  $\mathbf P(\mathsf X(t)=  t \mathsf v)$ are related, as in $d=1$,  to those of the partition function of a continuum directed polymers with noise variance of order $v^2= \Vert \mathsf v\Vert^2$. As a consequence,  it was predicted that for any $v\neq 0$ in $d=2$, and for 
$v>v_c$ in $d=3$, the fluctuations of $\log \mathbf P(\mathsf X(t)=t \mathsf v)$
scale as $t^{\beta}$ at large $t$, where $\beta$ is the DP 
exponent in dimension $d$. Furthermore, in $d=3$ for $v<v_c$ it
was predicted that the DP is in the weak noise phase
with bounded Gaussian EW type fluctuations. Finally $d=2$
being the critical dimension, disorder is marginally relevant
and from renormalization group arguments, the critical scale was predicted as $t \sim e^{C/v^2}$,
leading to a moderate deviation regime
in the scale $x \sim t/\sqrt{\log t}$. The recent preprint \cite{drillick2025random} confirms mathematically the order of the critical scale in $d=2$ and $d=3$ and describes the fluctuations in the sub-critical regime,  in terms of solutions to stochastic PDEs (heat equations with additive Gaussian noise). 
In particular, \cite{drillick2025random} consider the smoothed out field 
\begin{equation}
\mathbf I (\mathsf v, t) = \int  \varphi(x) C(\mathsf v, t, x) \mathbf P(\mathsf X(t)= t \mathsf v +t^{1/2} x)dx
\label{eq:defI}
\end{equation}
for   smooth functions $\varphi$ and some appropriate deterministic function $C(\mathsf v, t, x)$ (see \cite[Definition 1.10]{drillick2025random}).  The normalizing function is chosen so that  \be \overline{\mathbf I(\mathsf v, t)} \xrightarrow[t\to\infty]{} \int \varphi(x) \frac{e^{\frac{-x^2}{2}}}{\sqrt{2\pi}} dx.\ee 
In other terms, the normalizing constant is chosen so that $ \overline{C(\mathsf v, t, x) \mathbf P(\mathsf X(t)= t \mathsf v +t^{1/2} x)} $ converges to a heat kernel.  
Moreover, in reference \cite{drillick2025random}, the starting point $\mathsf X(0)$ is chosen to fluctuate on the $\sqrt{t}$-scale according to some probability distribution. 
Then, they prove that\footnote{the symbol $\asymp$ means that both sides are of the same order as $t\to\infty$, up to constants prefactors which, in the second line, may depend on $\mathsf v$.} 
\begin{equation}
    \mathrm{Var}( \mathbf I(\mathsf v, t)) \asymp \begin{cases}
       \frac{1}{t^{d/2}} & \text{ for } \vert\mathsf v\vert =O(t^{-1/2}) \\
       \frac{\vert \mathsf v\vert^2}{t^{\frac{d-2}{2}}} & \text{ for } \vert\mathsf v\vert < \mathsf v*
    \end{cases} \label{2regimes} 
\end{equation}
where the moderate deviation angle $v^*$ is  
\be 
\mathsf v^* =\begin{cases} t^{-1/4} &\text{ for }d=1,\\
    \frac{1}{\sqrt{\log t}} &\text{ for }d=2,\\
    \mathsf v_c  &\text{ for }d= 3.
\end{cases}
\ee 
For $d=3$, $\mathsf{v}^*=\mathsf{v}_c$ corresponds to the transition between strong disorder and weak disorder. 
For all dimensions, Ref. \cite{drillick2025random} shows that 
\be 
\frac{I(\mathsf v, t)- \overline{I(\mathsf v, t)}}{\sqrt{ \mathrm{Var}( \mathbf I(\mathsf v, t))  }}
\label{eq:convergenceGaussianDrillickParekh}
\ee 
converges to a Gaussian variable (characterized more precisely in \cite{drillick2025random} in terms of the solutions to a heat equation with additive noise). 

 We note an important difference between our work and the setting of \cite{drillick2025random}: in our work, we consider below the fluctuations of $\mathbf P(\mathsf X(t)= t\mathsf v)$ and $\log \mathbf P(\mathsf X(t)= t\mathsf v)$, not their smoothed out version, which obey different scaling behaviour.

Finally we mention the work  \cite{hass2025super} studying numerically the fluctuations of $\log \mathbf P(\Vert \mathsf X(t)\Vert \geq r)$ in dimension 
$d=2$. They find a variance of order $r^2/t^2$, which we note is consistent with \eqref{2regimes}  in both regimes $r \sim \sqrt{t}$ and $r \sim \mathsf  v t$.

\section{Dirichlet RWRE}
\subsection{Definition}
In higher dimensions $d>1$ a natural choice of environment distribution is the Dirichlet distribution  $\mathsf p\sim \mathrm{Dir}(\alpha_1, \dots,\alpha_{d+1}) = \mathrm{Dir}(\boldsymbol{\alpha})$ with probability density  
\be 
 \frac{1}{\mathcal N(\boldsymbol\alpha)} \prod_{i=1}^{d+1}  \left( p^{i} \right)^{\alpha_i}, 
\ee  
supported on $\sum_{i=1}^{d+1}p^i =1$,
where 
\be \alpha_{\circ}  =\sum_{i=1}^{d+1}\alpha_i, \;\;\; 
\mathcal N(\boldsymbol\alpha) = \frac{\prod_{i=1}^{d+1} \Gamma(\alpha_i)}{\Gamma\left(\alpha_{\circ}\right)}.
\ee 
When $d=1$, we recover the Beta random walk model, which is Bethe ansatz solvable \cite{barraquand2017random}.

The annealed random walk has drift $\mathsf v_\circ$ with coordinates 
\be
v_{0,i}= \frac{\alpha_i}{\alpha_\circ} - \frac{1}{d+1} 
\ee 
($i=1,\dots,d+1$) and diffusion matrix with entries
\be 
\mathsf D_{ij}= \delta_{i j} \frac{\alpha_i}{\alpha_\circ}
- \frac{\alpha_i \alpha_j}{\alpha_\circ^2}.
\label{eq:defDDirichlet}
\ee 
One uses that $p_i \sim \mathrm{Beta}(\alpha_i, \alpha_\circ-\alpha_i)$,
leading to $\overline{p_i}= \alpha_i/\alpha_\circ$.
In the case $\alpha_i=\alpha$ , the drift vanishes $\mathsf v_0=0$ and one has 
\be 
\mathsf D_{ij}= \frac{1}{d+1} \delta_{i j}  
- \frac{1}{(d+1)^2}  \quad , \quad D = \frac{d}{d+1} 
\ee
\subsection{Stationary measure and local central limit theorems}
For any dimension, the Dirichlet RWRE has an explicit stationary measure, in the following sense. Define a dynamics on measures $\mu_t$ by 
\be 
\mu_{t+1}(x) = \sum_{y} \mathbf P\left( \mathsf X(t+1)=x \vert \mathsf X(t)=y\right) \mu_t(y). 
\ee 
If at some time $t$, the random variables $\mu_t(x)$ are independent Gamma distributed with parameter $\alpha_{\circ}$, then $\mu_{t+1}$ has the same distribution. 

In the case $d=1$, we recover a result from \cite{thiery2016exact,barraquand2023random}. When $x=O(\sqrt t)$, in the large time limit, 
\be 
\mathbf P(\mathsf X(t)=\mathsf v_\circ t + x)\sim \frac{e^{\frac{-x^2}{2Dt}}}{\sqrt{2\pi D t}} \frac{\gamma_x}{\alpha_{\circ}}
\ee 
where $\gamma_x$ are independent  Gamma random variables with parameter $\alpha_{\circ}$.  

The general form of the stationary measure in dimension $d$ relies on the following two properties: \begin{enumerate}
    \item if $\gamma\sim \mathrm{Gamma}(\alpha_{\circ})$ and $\mathsf p\sim\mathrm{Dir}(\alpha_1, \dots, \alpha_{d+1})$ are independent, then $(p_1\gamma, \dots, p_{d+1}\gamma)$ are independent Gamma variables with parameters $\alpha_1, \dots , \alpha_{d+1}$ \cite{mosimann1962compound}.
\item the sum of independent gamma variables with parameters $\alpha_i$ is a Gamma random variable with parameter $\alpha_{\circ}$. 
\end{enumerate}  
Moreover in general dimension $d$, the random walk satisfies a local central limit theorem \cite{boldrighini1997almost}. Combined with the knowledge of the stationary measure, this implies that when $\mathsf x=O(\sqrt{t})$ and $t$ goes to infinity, 
\be 
\mathbf P(\mathsf X(t)= \mathsf x) \sim \frac{e^{- \frac{\mathsf x^T \mathsf D^{-1} \mathsf x}{2t  }}}{(2\pi t)^{d/2}\sqrt{\widehat{\det} \mathsf D} }\frac{\gamma_{ \mathsf x}}{\alpha_{\circ}}, 
\label{eq:localCLT}
\ee 
where $\gamma_{\mathsf x}$ are again independent  Gamma random variables with parameter $\alpha_{\circ}$.  
Since the $(d+1)\times(d+1)$ matrix $\mathsf D$ defined in \eqref{eq:defDDirichlet} has rank $d$, one should understand $\mathsf D^{-1}\mathsf x$ as a generalized inverse, well-defined on the hyperplane $x\in \mathbb T_d$, and $\widehat{\det}(\mathsf D)$ is defined as the product of all non-zero eigenvalues of $\mathsf D$.

As a consequence, for $\mathsf n = t \mathsf d + O(\sqrt{t})$, the $k$-th cumulant for $k \geq 2$
\be \label{cumulants} 
\kappa_k\left( \log  P(\mathsf n) \right) \xrightarrow[t\to\infty]{} \psi^{(k-1)}(\alpha_{\circ}).
\ee 
where $\psi(z)$ is the digamma function.
In particular, 
\be 
\Var \log  P(\mathsf n) \xrightarrow[t\to\infty]{} \psi'(\alpha_\circ). 
\label{eq:limitVariancetypical}
\ee
Indeed, if $w=\log \gamma$, the cumulant generating function of $w$ reads 
\be 
\langle e^{s w} \rangle = \frac{\Gamma(s+\alpha_{\circ})}{\Gamma(\alpha_{\circ}) } 
\ee

\subsection{Connections to directed waves}
The Dirichlet RWRE is closely connected to a generalization of a  model of directed waves in random media introduced in \cite{saul1992directed}. We
introduce in Appendix \ref{sec:SKR} a wavefunction $\Psi(\mathsf n)\in \mathbb C^{d+1}$ defined for $\mathsf n\in \Z^{d+1}$ such that $\Vert\Psi(\mathsf n)\Vert^2=P(\mathsf n)$ for all $\mathsf n\in \mathbb Z^{d+1}$. The wavefunction 
$\Psi(\mathsf n)$ undergoes a unitary evolution determined by independent Haar distributed random matrices in $U(d+1)$ attached to each site $\mathsf n$. The Dirichlet distribution with parameter $\alpha_1=\dots=\alpha_{d+1}=1$ arises because this is the law of the moduli squared of matrix elements along a column for random Haar unitary matrices.

\section{Scaling of the variance and KPZ growth exponent} 
\label{sec:variance}
For simplicity,  we  restrict from now on to the case $\alpha_1=\dots=\alpha_{d+1}=\alpha$. 
In this section, we consider the variance ${\rm Var} \log P(\mathsf n)$ in three possible types of directions, in dimensions $d=1,2,3$. 
\begin{enumerate}
    \item Typical directions ($\mathsf n/t\approx  \mathsf d=(\frac{1}{d+1},\dots,\frac{1}{d+1})$), i.e.,along the diagonal.
    \item Generic atypical directions ($\mathsf n=t\mathsf u$, $\mathsf u\neq\mathsf d$). The vector $\mathsf u\in (\mathbb R_+)^{d+1}$ is chosen so that $u_1+\dots+ u_{d+1}=1$, $\mathsf u\neq \mathsf d$. We also assume that  $\mathsf u$ is not be on the boundary of the simplex (that is $u_i\in (0,1)$ for all $1\leq i\leq d+1$ but it cannot be $0$ or $1$)
    \item Boundary cases. At least one of the $u_i$ is zero.
\end{enumerate}
Concretely, we may choose vectors $\mathsf u$ of the form
\be 
\mathsf u=\left( \frac{k-d}{k}, \frac{1}{k}, \dots, \frac{1}{k} \right)
\label{eq:exampleu}
\ee 
parameterized by some $k\in [d, +\infty]$. The typical direction corresponds to $k=d+1$. The extreme cases $k=d$ and $k=+\infty$ are boundary cases, which, for $d\geq 2$ behave differently. 

\subsection{Theoretical prediction}
In typical directions, we expect to see the variance of the stationary measure, 
\be 
\Var \log P(\mathsf n) \xrightarrow[t\to\infty]{} \psi'((d+1) \alpha). 
\label{eq:limitVariancetypical2}
\ee

In generic atypical directions, we expect that at large $t$ the variance scales as 
the KPZ class fluctuations. 
Let $\mathsf u\in (\mathbb R_+)^{d+1}$ be a fixed vector chosen as above. 
We expect that 
\begin{equation}
      \Var [ \log P(t \mathsf u ) ] \simeq c(\mathsf u, \alpha) t^{2 \beta}  
      \label{eq:predictionvariancelogP}
\end{equation}
where $\beta$ should be the universal exponent in $1+d$ KPZ growth and
$c(\mathsf u,\alpha)$ is a constant (of order $1$ as $t$ goes to infinity). 
The KPZ growth exponent is known exactly for $d=1$ with 
$\beta_{1+1}=1/3$, and from numerical simulations 
in $d=2,3$ with $\beta_{2+1}=0.2398$ (extracted from \cite{pagnani2015numerical}
which found $\chi_{2+1}=0.3869(4)$) and $\beta_{3+1} = 0.184$
\cite{odor2010directed}. 
In dimension $d=3$,  we expect \cite{ledoussal2017diffusion} that  there is a transition, i.e.\ $c(\mathsf u, \alpha)$ 
may vanish when the vector $\mathsf u$ is closer to the typical direction
than some threshold value, which corresponds to the weak disorder 
phase of the KPZ equation. 
In that phase, 
the variance remains finite at large $t$. 

In the boundary cases, when the direction $\mathsf u$ is chosen so that some of the $u_i$ equal zero, the variance of $\log P(\mathsf n)$ is expected to grow as a power law of time with exponent $\beta_{1+d'}$ where $1+d'$ equals the number of nonzero coordinates in $\mathsf u$. 
For instance, if $\mathsf u$ is chosen as in  \eqref{eq:exampleu}, for $k=+\infty$, $\mathsf u=\mathsf e_1$, all steps of the walk  are in the $\sf e_1$ direction and $\log P(t \mathsf e_1)$ follows the central limit theorem and its fluctuations grow like $t^{1/2}$. 
More precisely, for the variance,
\be  
\Var \log P(\mathsf n) = t \Var \log p^1 = t \, \left( \psi'(\alpha)-\psi'((d+1)\alpha) \right)
\label{eq:variance:edge}
\ee
If $k=d$, $\mathsf u=(0,\frac{1}{d}, \dots, \frac{1}{d})$, we expect that 
\be 
\Var \log P(\mathsf n) \simeq c_{d'}(\mathsf u, \alpha) t^{2\beta_{1+d'}},\;\;  d'=d-1.
\label{eq:predictionsubspace}
\ee 
In this case, the random walk is confined in a $(1+d')$-dimensional subspace and $\log P(\mathsf n)$ can be directly identified with the free energy of a directed polymer model in dimension $1+d'$, with random weights $p^2_{\mathsf n,t} ,\dots, p^{d+1}_{\mathsf n,t}$. We stress that here, there is no distinction between typical and atypical directions in the subspace.
For instance, for $d=2$, the random weights $p^2$ and $p^3$ are such that $p^2+p^3= 1-p^1<1$ is random. Moreover, we expect that in that case after rescaling, the probability distribution of $\log P(\mathsf n)$ converges to the Tracy-Widom distribution.

\subsection{Numerical implementation}
We have checked these predictions by computer simulations. To numerically \cite{practical_guide2015} compute the probabilities, 
one could store in the computer code the data in $(d+1)$-dimensional arrays $P$ at positions $\mathsf n$, e.g. entries $P[n_1][n_2]$ for $d=1$. For any time $t$ we now call those sites $\mathsf n$ \emph{accessible}  where $\sum_i n_i=t$.
Within the algorithm, one has to iterate over all accessible sites for  increasing times $t=0,1,2,\ldots$ and perform some \emph{action}. These \emph{actions} can be
initializing the probabilities to zero or calculating the contributions for the next time $t+1$. This could be implemented in the code 
for $d=1$ by a loop {\bf for}($n_1=0 ... n_1=t$), letting $n_2=t-n_1$ and each time perform the \emph{action}. For $d=2$ one would need a three-dimensional array to store the values of $P$, with two nested loops and in $d=3$ one would need a four-dimensional array and three nested loops. This would be quite tedious and not very flexible since a different code would be needed for every considered value of $d$. Also, the arrays would consume quite a bit of memory if $d$ and $t$ are large.

Thus instead we store the probabilities in vectors $\vec P_t$, 
i.e.\  arrays,
for each time $t$. The numbers of entries of $\vec P_t$ is equal to the number
of accessible sites, that is $a_{1,t}\equiv t+1$ entries for $d=1$,
$a_{2,t}\equiv (t+2)(t+1)/2$ for $d=2$ and $a_{3,t}\equiv (t+3)(t+2)(t+1)/6$ for $d=3$.
Technically $P_t$ is a one-dimensional array, and we use the values of $a_{d,t}$ to calculate for any value of $\mathsf n$ (with $\sum_i n_i=t$) the position within the array. 

In order to reach 
large times $t=0,\ldots,t_{\max}$, we only hold $\vec P_t$ and $\vec P_{t+1}$
in memory, starting at $t=0$ and $P(\mathsf 0)=1$.  
At each time $t$ we
iterate over all accessible sites $\mathsf n$ and assign the $d+1$ "outgoing" probabilities
$\mathsf{p}_{\mathsf n,t}$ from the Dirichlet distribution. Here, we use the
\emph{GNU Scientific Library} to sample the corresponding random numbers. We use  these numbers to add up the $d+1$ contributions $p^i_{\mathsf n,t}P(\mathsf n)$ to the values of $P(\mathsf n+\mathsf e_i)$. To store in our C program the actual probabilities $P(\mathsf n)$, which can become very small like 
$10^{-1000}$, we use a custom data type of two {\tt double} values, one for the mantissa, one for the exponent, plus the corresponding arithmetic operations.

To conveniently iterate over all accessible sites,  i.e.,exploring all points of the discrete simplex one by one without explicitly using nested loops, we use a recursive approach, which works the same way for all values of $d$ and is given by the following algorithm. 

\begin{tabbing} xx \= xx \= xx \= xx \= xx \= xx \= xx \kill
{\bf algorithm} iterate($\mathsf n$, $d'$, $t'$)\\
{\bf begin} \\
\> {\bf if $d'>0$} {\bf then}\\
\>\> {\bf for} $k=0$ .. $t'$ {\bf do} \\
\>\>\>  $n[d']=k$\\
\>\>\> iterate( $\mathsf n$, $d'-1$, $t'-n[d']$)\\
\>\> {\bf done}\\
\> {\bf else}\\
\> \> $n[0]=t'$\\
\> \> perform \emph{action} at $\mathsf n$\\
{\bf end}\\
\end{tabbing}

The vector $\mathsf n$ is used to store iteratively the current
accessible site one after the other,
thus it is set during the recursive execution of the function. 
The value of $d'$ states how many entries of $\mathsf n$ have to be assigned, so for the initial call it is $d'=d+1$. The value of $t'$ states how many "time units" have to be given to these remaining $d'$ entries, i.e.,they have to sum up to $t'$. Therefore, for the initial call $t'=t$ is used, i.e., the current time.
For each accessible site, which is obtained at the deepest level $d'=0$ of the recursion, the \emph{action} is performed, e.g., an update of the probabilities. Note that due to the recursion an implicit nesting of loops occurs.

\subsection{Numerical results}

\begin{figure}
\begin{center}
\includegraphics[width=0.35\textwidth]{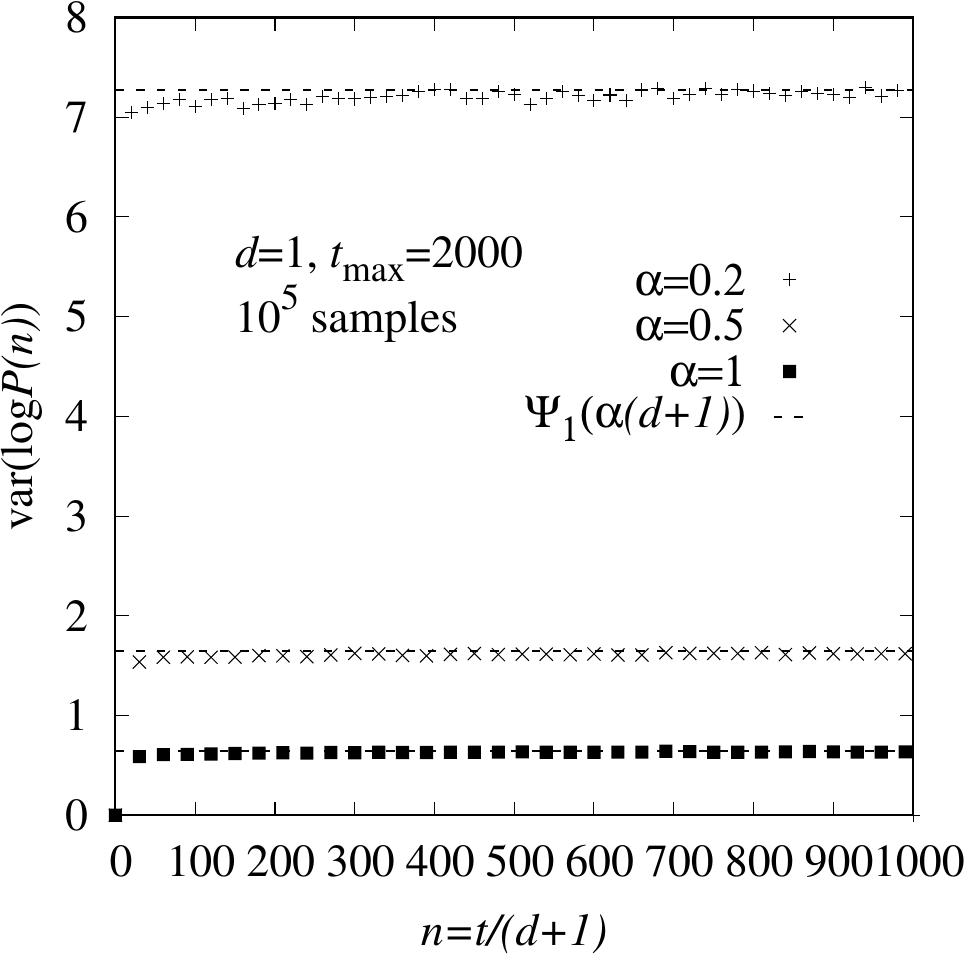}
\includegraphics[width=0.35\textwidth]{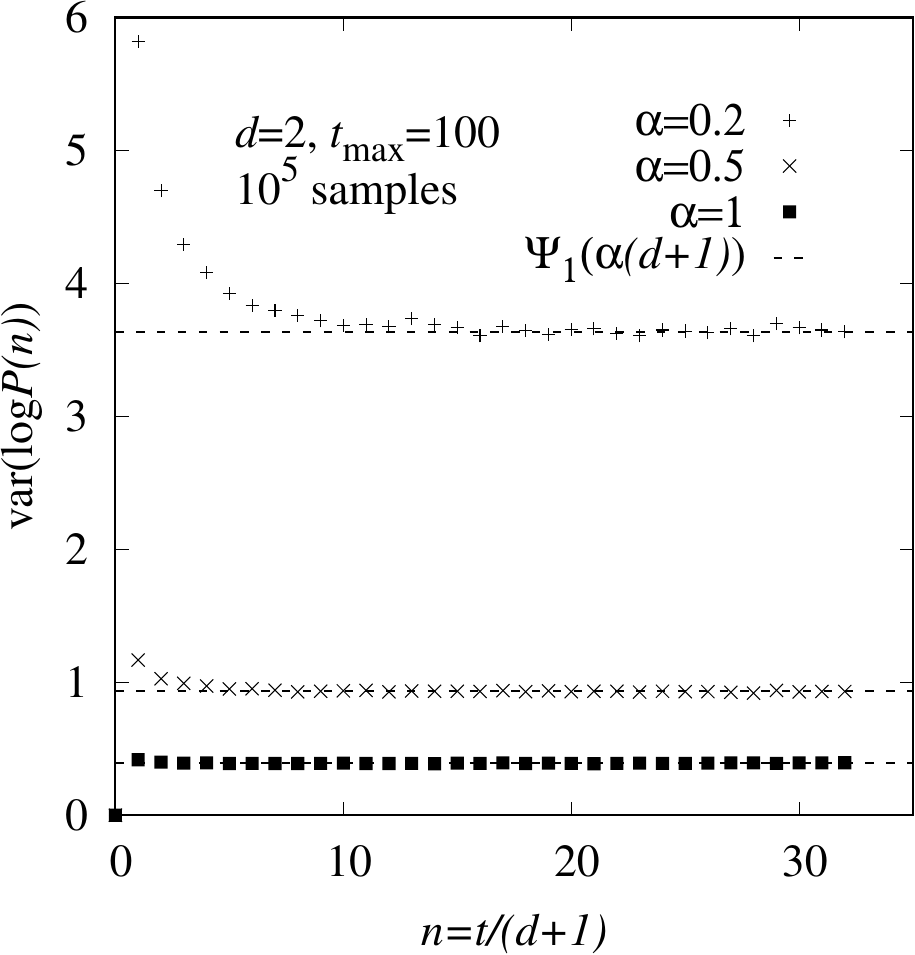}
\includegraphics[width=0.35\textwidth]{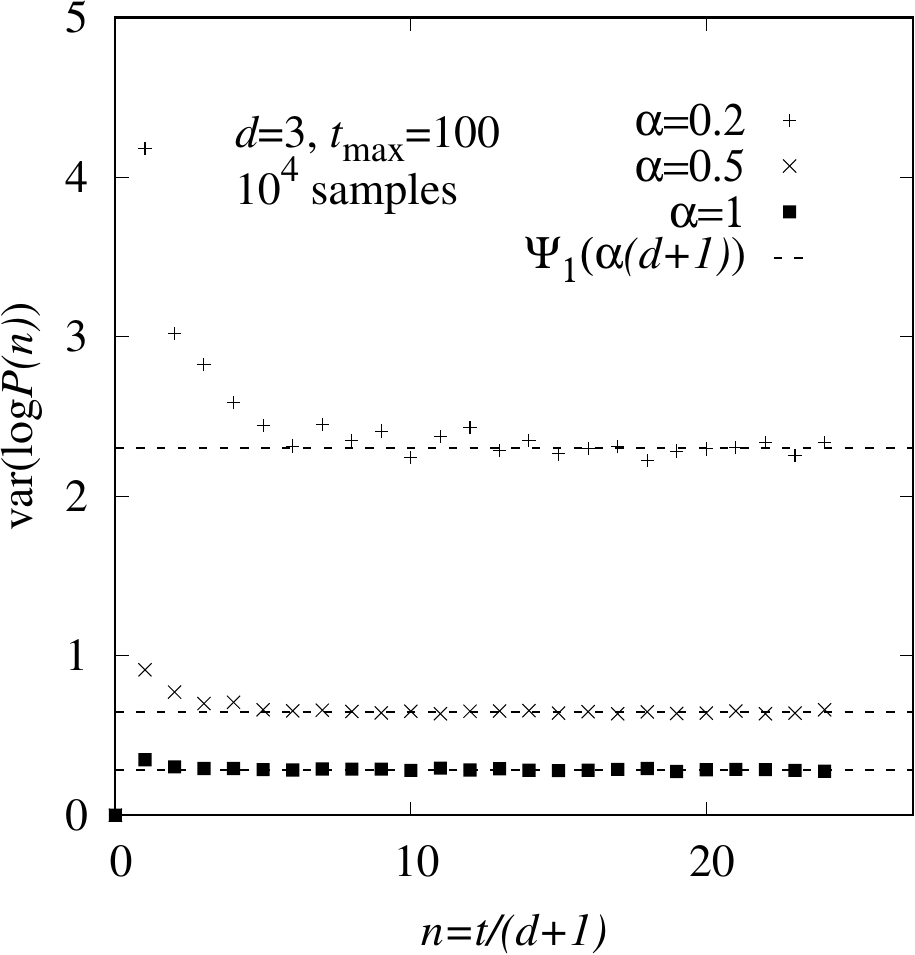}
\end{center}
\caption{\label{fig:varLogP_d2} 
The numerically estimated variance of the log of the probability in the typical direction for $d=1$ (top, $\mathsf u=(1/2,1/2)$), $d=2$ (middle, $\mathsf u=(1/3,1/3, 1/3)$) 
and $d=3$ (bottom, $\mathsf u=(1/4,1/4,1/4, 1/4)$ as a function of the time $t$,
for different values of $\alpha$. The estimates were performed by averaging over $10^5$ ($d=1$, $d=2$) or $10^4$ ($d=3$) samples of the environment, respectively. A fast convergence towards the values given in
Eq.~(\ref{eq:limitVariancetypical2}), as displayed by horizontal lines, is visible.}
\end{figure}

The predictions in the typical direction are verified in dimensions $d=1,2,3$, see Figure \ref{fig:varLogP_d2}. The convergence in time $t$ to the predicted asymptotic value 
\eqref{eq:limitVariancetypical2}
is quite
fast. 

We now turn to the other directions. 
\subsubsection{Dimension $d=1$}
 
\begin{figure}
\begin{center}
\includegraphics[width=0.4\textwidth]{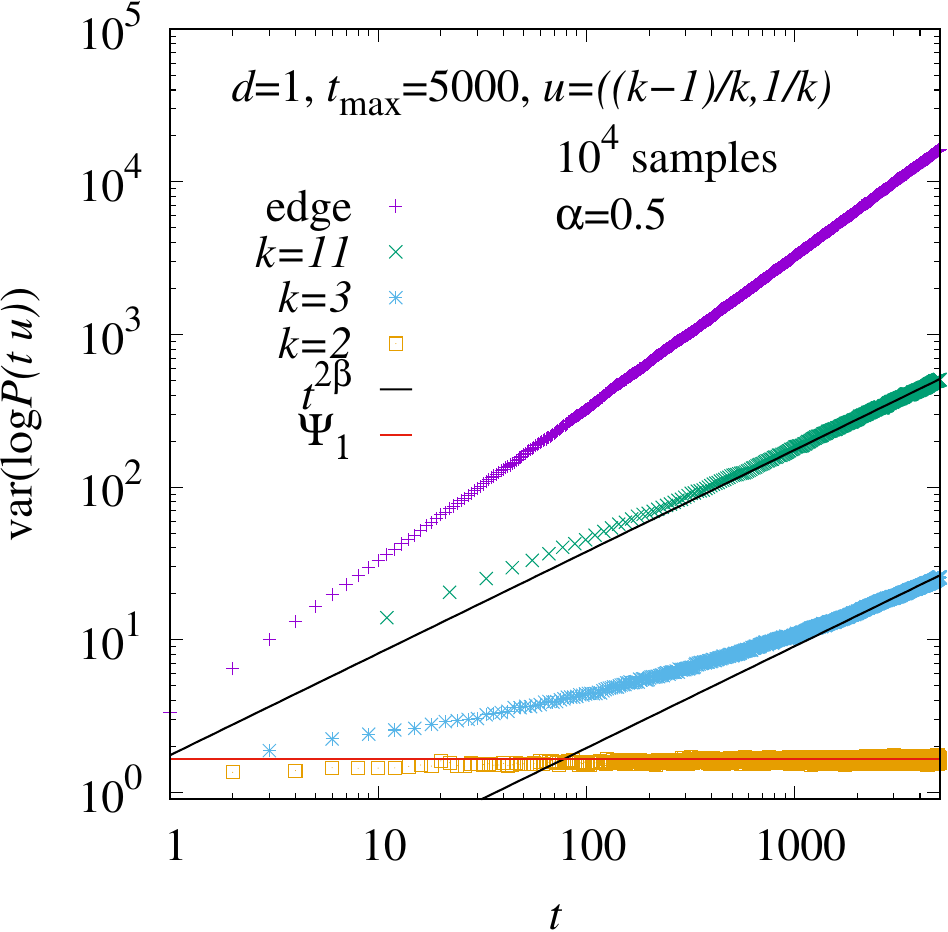}
\end{center}
\caption{\label{fig:varLogP_k_d1} 
The variance of the log of the probability as function of time $t$ for $d=1$ and $\alpha=0.5$, measured along
several directions $\mathsf u$. The solid lines represents the expected asymptotic behaviour $t^{2/3}$.}
\end{figure}

For $d=1$ results are shown in Fig.~\ref{fig:varLogP_k_d1} for 
four vectors:  the diagonal $\mathsf u =\mathsf d = (1/2,1/2)$, off diagonal
vectors $\mathsf u=(2/3, 1/3)$, $\mathsf u = (10/11,1/11)$ and an edge $(1,0)$. 
The results were obtained for $\alpha=0.5$, a maximum time $t_{\max}=5000$
and averaged over $10^4$ samples.
 For the diagonal, the above described convergence to a constant is visible. For the two
off diagonal cases, in particular $\mathsf u=(10/11,1/11)$, 
a power law behavior is visible for large values of $t$, which is compatible with $t^{2\beta}$, where $\beta=1/3$ is the KPZ growth exponent \cite{forster1977large, huse1985huse, kardar1986, baik1999distribution} for $d+1=1+1$
dimensions. For an edge
vector $\mathsf u=(0,1)$ a different power law is visible, according
to Eq.~(\ref{eq:variance:edge}) we have
$\log P(t\mathsf u) = (\psi'(1/2)-\psi'(1)) t \approx 3.29 t$.
A fit of the data for $t\ge 1000$ results a prefactor of $3.35(4)$ which
is compatible with this prediction.
We have also performed simulations for $\alpha=0.2$ and $\alpha=1$
which lead to similar results.

\begin{figure} 
\begin{center}
\includegraphics[width=0.4\textwidth]{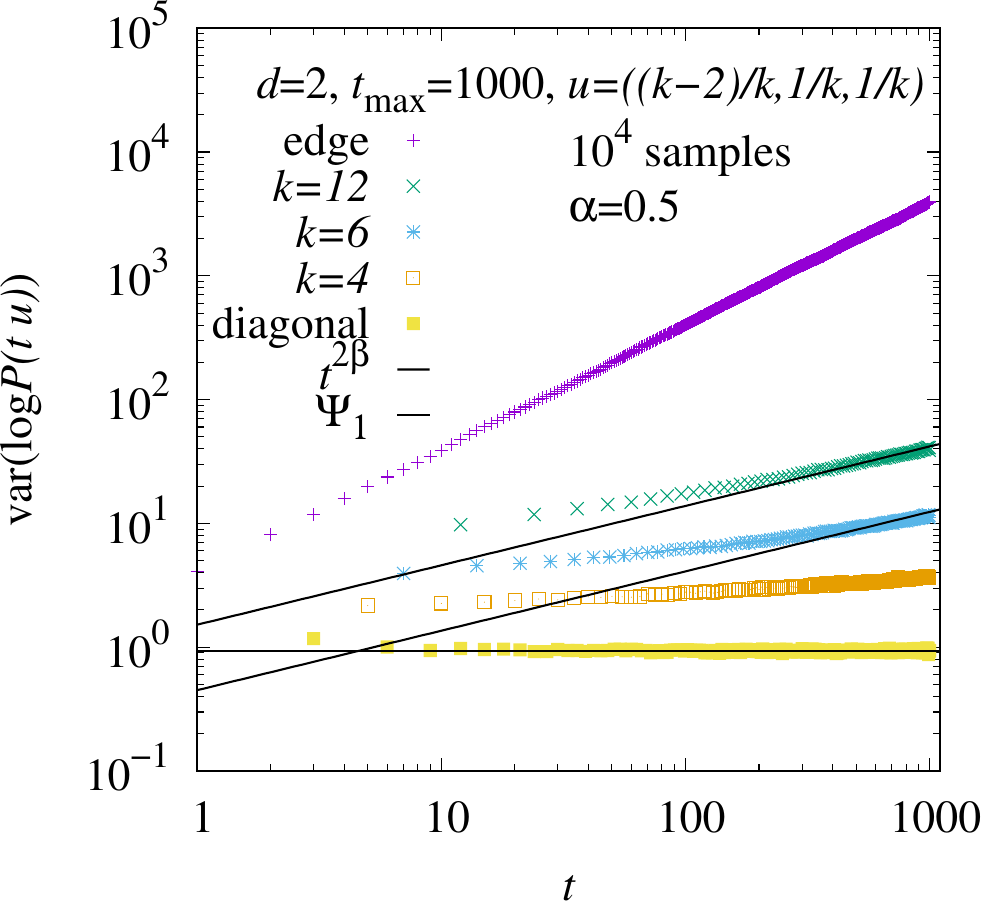}
\end{center}
\caption{\label{fig:varLogP_k_d2} 
The variance of the log of the probability as function of time $t$ measured along
several directions $\mathsf u$, for $d=2$ and $\alpha=0.5$. The solid
ascending lines correspond to
$t^{2 \beta}$ with $\beta=0.24$,
while the horizontal lines shows $\Psi_1(\alpha(d+1))$.
}
\end{figure}

\subsubsection{Dimension $d=2$}
The corresponding results for $d=2$ and $\alpha=0.5$, with $t_{\max} =1000$ 
and with an average over $10^4$ damples are shown in
Fig.~\ref{fig:varLogP_k_d2}. Here the results for the vectors $\mathsf u = (1/3,1/3,1/3)$,
$\mathsf u=(2/4, 1/4, 1/4)$, $\mathsf u=(4/6,1/6,1/6)$, $\mathsf{u}=(10/12,1/12,1/12)$
and the edge $\mathsf u =(1,0,0)$ are shown. 
For off-diagonal vectors, close to the edge, a power law behavior
can be observed which is compatible with $\sim t^{2\beta}$  for $\beta=0.2398$
which was observed numerically \cite{marinari2000,pagnani2015numerical} 
for $d+1=2+1$ dimensional KPZ. Since for $d=2$
we cannot reach as large times $t$ as for $d=1$, our results cannot
rule out a different power-law behavior. For the edge case, $\mathsf u=(1,0,0)$, 
again a linear behavior is visible, here the data for large values of $t$
is compatible with $4t$, which agrees with
the prediction Eq.~(\ref{eq:variance:edge}) since $(\psi'(1/2)-\psi'(3/2))=4$.

\begin{figure}
    \centering
    \includegraphics[width=0.4\textwidth]{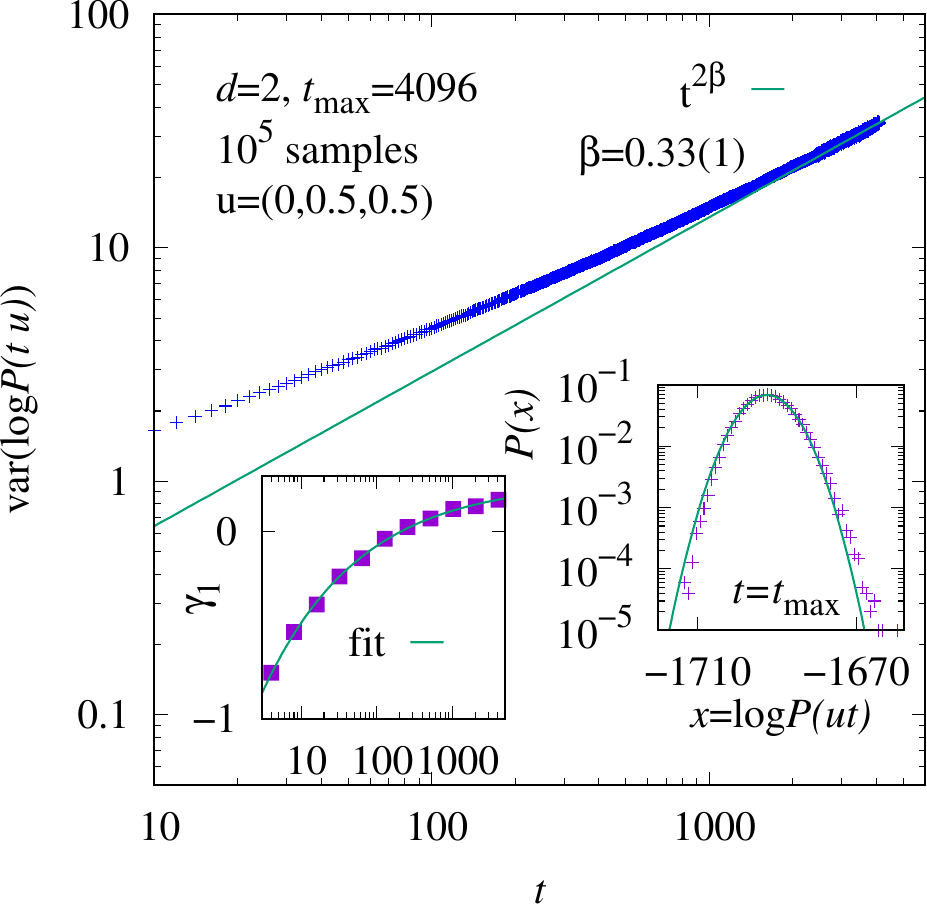}
    \caption{Main plot: The variance of the log of the probability
      as function of time $t$ in direction $\mathsf u=(1/2,1/2,0)$. The line
      indicates the power law $t^{2\beta}$ with $\beta=1/3$.
      Left inset: skewness of the distribution again as function of time.
      The line shows a fit to a power-law plus constant
      Eq.~(\ref{eq:power:correction}) Right inset:
      distribution at the largest time $t_{\max}=4096$.
      The line shows a corresponding
      Gaussian, fitted to the peak region. One sees that all off-peak data points are
      systematically below of the Gaussian left of
      the peak, and above the Gaussian right of the peak.
    \label{fig:subspace}}
\end{figure}

For the edge vector $\mathsf u=(0,1/2,1/2)$ the data in Fig.~\ref{fig:subspace} shows that 
at large time
\be 
\Var \log (P(\mathsf n)) \simeq c_1 t^{2/3}\,,
\ee 
which is again compatible with the KPZ exponent $\beta=1/3$
for the corresponding 1+1-dimensional subspace.
We expect that $\log (P(\mathsf n))$ converges to the
TW distribution. At least we are able
to see that the distribution, for increasing time $t$, becomes
slightly tilted to the right as compared to the Gaussian.
This is directly visible in the right
inset of Fig.~\ref{fig:subspace}.
We have also obtained the 
third cumulant, which for a random variable $X$ reads
\begin{equation}
    \kappa_3= 
    \overline{ ({X-\mu})^3} =
    \overline{X^3} - 3\mu \sigma^2-\mu^3
    \label{eq:third:cumulant}
\end{equation}
and the skewness
\begin{equation}
    \gamma_1 = \kappa_3/\sigma^3\,. 
    \label{eq:skewness}
\end{equation}
Recall that $\overline{\rule{0pt}{2.5mm}\cdots}$ denotes the expectation with respect to
the disorder samples, $\mu=\overline{X}$ and 
$\sigma^2$ is the variance $\overline{X^2}-\mu^2$. Here we have 
$X=\log P(t\mathsf u)$. The time development of the skewness is shown
in the left inset of \ref{fig:subspace}. Here we have fitted the time development
to a power-law function with correction term of the form
\begin{equation}
\gamma_1(t)=\gamma_1^{\infty}+
\tilde b t^{-\tilde c}(1+\tilde d t^{-\tilde e})\,,
\label{eq:power:correction}
\end{equation}
yielding $\gamma_1^{\infty}=0.26(10)$, which is due to the rather
large error bar compatible with the skewness 0.22 of the $F_2$
TW distribution.
For a better verification of the compatibility with a TW
distribution, a larger time $t_{\max}$,
many more samples or maybe even a large-deviation
approach \cite{kpz2018,hartmann2024probing,les_houches2025}
would be necessary.

\subsubsection{Dimension $d=3$}
\begin{figure}
\begin{center}
\includegraphics[width=0.4\textwidth]{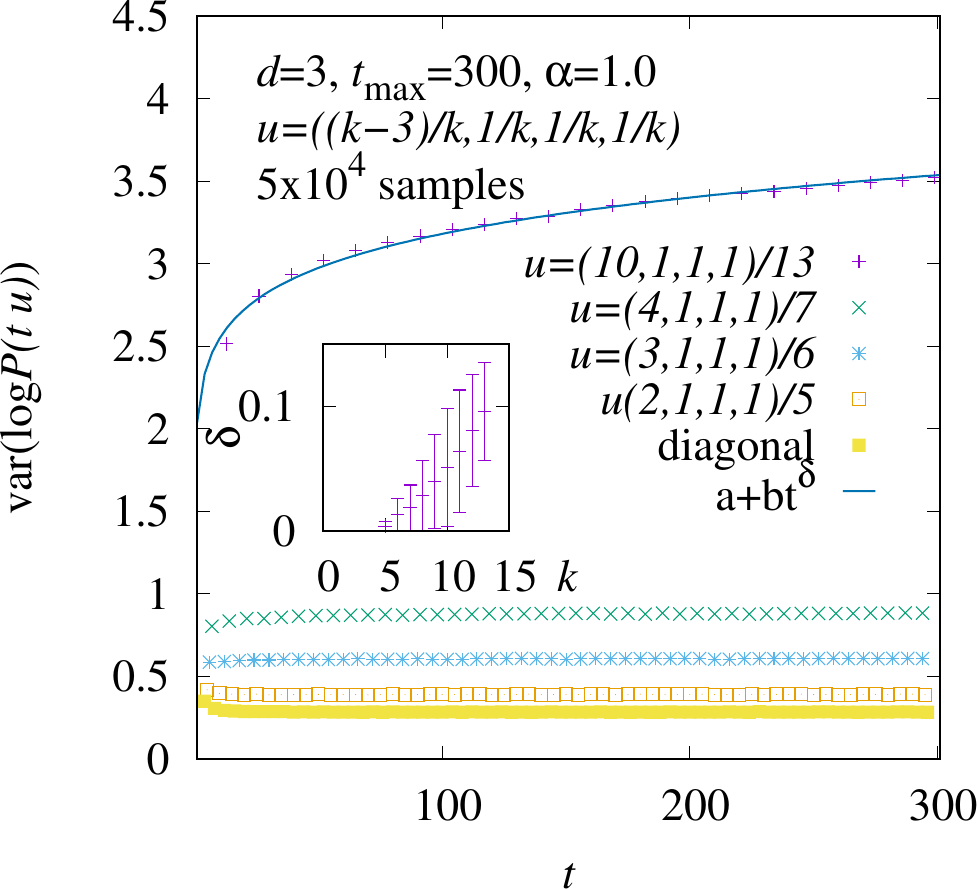}
\includegraphics[width=0.4\textwidth]{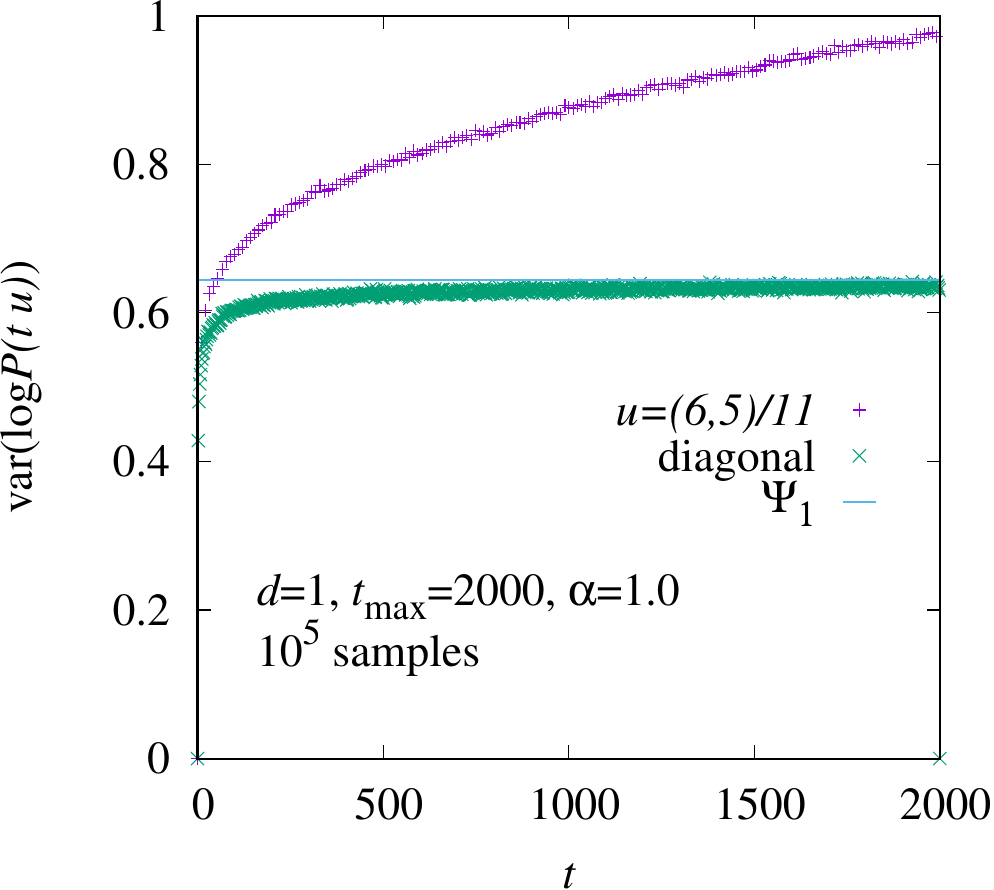}
\end{center}
\caption{\label{fig:varLogP_k_d3} 
Top: The variance of the log of the probability as function of time $t$ measured along
certain directions $\mathsf u(k)=(k-3,1,1,1)/k$ for 
$d=3$. 
The inset in the right shows results for the exponent $\delta$ when fitting the $d=3$ data to power laws of the form
$a+bt^{\delta}$. 
For small values of $k$, the exponent is within
a one-sigma error bar compatible with 0, while it is not for large $k$. \\
Bottom: For comparison, we consider the same quantity in dimension $d=1$. In that case, the off-diagonal growth in time is well visible, even for relatively small angles.}
\end{figure}

For $d=3$ it is expected that there is a range of vectors $\mathsf u$ close
to the diagonal, where var(log($P(\mathsf u t)$) does not increase with time $t$.
Our results are shown in Fig.~\ref{fig:varLogP_k_d3}
for vectors $\mathsf u(k)=(k-3,1,1,1)/k$ ($k>3$). One observes that for a small angle, $\mathsf u=(2,1,1,1)/5$, the variance initially decreases slightly and then appears
to remain constant. For a slightly larger angle,  $\mathsf u=(3,1,1,1)/6$,  the data is compatible with  a convergence to a constant value. However, for the vector $\mathsf u=(10,1,1,1)/13$, which is much different from the diagonal, an increase in time is visible.

For further check, we have fitted power-laws plus a constant
of the form $a+bt^{\delta}$ to the data, where we have restricted $a>0$ and $b>0$. Here the exponent $\delta$ shows an increase
as a function of $k$, see inset in the top of Fig.~\ref{fig:varLogP_k_d3}. Note that for $k<10$ the
one-sigma error bars allows for a value $\delta=0$ but not
for larger value of $k$. This speaks in favor of a change
from a limiting behavior for small $k$ to a growth for larger $k$.

Thus, the results are compatible with a strong difference between close-to typical and very atypical directions, for $d=3$ but not for $d=1$ shown in Fig. \ref{fig:varLogP_k_d3} bottom, or $d=2$  (not shown here) as predicted. 
In dimension  $d=3$,  we are limited in the maximum times which can be achieved, since each runs takes a time $O(t_{\max}^4)$ and an average over many runs is need. Thus, $t_{\max}=1000$ and an average over $10^4$ samples would
need of the order of $10^{16}$ operations. Thus, in contrast to $d=2$ and $d=1$, we are not able to verify the power-law behavior of the variance 
for larger times \eqref{eq:predictionvariancelogP}
with the expected KPZ exponent $2 \beta_{3+1} \approx 0.37$.

\section{Phase transition in dimension 3}

\subsection{Theoretical predictions}

It is known that for directed polymers on $\Z^{d+1}$ at inverse temperature $\beta$, for $d\geq 3$, the polymers obey a transition between weak disorder for $\beta\in (0, \beta_c)$ and strong disorder for $\beta>\beta_c$. Moreover, the partition function $Z_n(\beta)$ associated to paths of length $n$ is such that for $\lambda(\beta) = \log \overline{Z_n(\beta)}$, the second moment 
$\overline{e^{-2\lambda(\beta)}  Z_n(\beta)^2 }$ remains bounded as $n$ goes to infinity for $\beta\in [0, \beta_2)$,  for some critical value $\beta_2 \leq \beta_c$ \cite{junk2024strong, junk2025coincidence}, and it diverges when $\beta\geq \beta_2$. 

Likewise, we expect a similar transition for random walks in random environment as a function of the velocity $v$. We denote $v_c$ the critical velocity. 
Let us introduce the random variable 
\be 
{\cal Z}_t = \frac{\mathbf P(\mathsf X(t)=t v \mathsf s_1) }{\overline{\mathbf P(\mathsf X(t)=t v \mathsf s_1)}}
\ee 
where we have chosen $\mathsf{v}=v \mathsf s_1$.
We expect that in the weak noise phase $(v\leq v_c)$ its distribution reaches a limit distribution ${\cal P}({\cal Z})$, with, for $v>0$, a power law heavy tail
\be \label{heavytail} 
{\cal P}({\cal Z}) \sim \frac{1}{{\cal Z}^{1+ \mu}} \quad , \quad \mu=\mu(v)
\ee 
at large ${\cal Z}$, where the exponent $\mu$ depends on $v$ and diverges as $v \to 0$. Indeed this is what is found for the rescaled partition function of directed polymers in weak disorder \cite{junk2024tail}, see also \cite{meerson2018nonequilibrium}. 

As a consequence, we expect the $p$-th moment $\overline{(\mathcal Z_t)^p}$ to converge as $t$ to infinity to a finite value for $p<\mu(v)$. In the strong disorder phase $(v>v_c)$, all moments $\overline{(\mathcal Z_t)^p}$ diverge to infinity at large time. 

It is instructive to consider the second moment of $\mathcal Z_t$, which we can compute exactly using two replicas of the random walk. The calculation is performed in Appendix \ref{sec:secondmoment}. The result takes the form 
\be \label{secondmoment}
\lim_{t\to\infty} \overline{\mathcal Z_t^2} 
 = \frac{ 1 + \frac{1}{4 \alpha}}{(1- \frac{3v^2}{4 \alpha}  L(a))^2 } .
\ee 
where $e^a = (1+ 3 v)/(1-v)$ and  $L(a)$ has the explicit expression  given in \eqref{La}. The quantity $L(a)$ corresponds to the total local time of intersection of two annealed random walks which are biased so that they escape ballistically as $t v\mathsf s_1$.
The divergence in \eqref{secondmoment} corresponds to the formation of a 2 replica 
bound state.

For vectors of the form $\mathsf u(k) = (\frac{k-3}{k}, \frac{1}{k},\frac{1}{k},\frac{1}{k}) $, we have the relation $v=1-4/k$, and we find (when $\alpha=1$) the table of numerical values for $m_2(k) = \lim_{t\to\infty} \overline{(\mathcal Z_t)^2}$
shown in Tab.~\ref{tab:2nd:moment}.

We find that there is a threshold velocity $v=v_2$, and a corresponding value $k=k_2$, such that $m_2(k)$ is bounded for $v\in (0, v_2)$, and diverges to infinity for $v\geq v_2$. We estimate the value of $v_2 \approx 0.639$, which corresponds to a threshold value of $k$ being $k_2\approx 11.1$.
Thus it implies the lower bound on the critical velocity
\be 
v_c \geq 0.639,  \quad  \quad k_c \geq 11.1 
\ee 

It would be interesting to study the index $\mu(v)$ 
as $v$ varies.  
From the study of the analogous question for the partition function of directed polymers \cite{junk2024tail} (see also 
\cite{lacoin2025localization,junk2024strong,junk2025coincidence}),  
we expect that this limit distribution has a power law tail 
with exponent $ \mu(v_c) = 1+ 2/d$, i.e., $\mu(v_c)=5/3$ when $d=3$.

Next, we consider the  third cumulant, because we expect that there exist a critical angle $\theta_{\rm c}$ such  that for angles $\theta=\angle(\mathsf d, \mathsf u)$ between the diagonal $\mathsf d$ and vector $\mathsf u$ the behavior changes
at $\theta_{\rm c}$: 
\begin{equation} \label{predictionlogP}
    \log P(t\mathsf u) \simeq \begin{cases}
        \log \xi_{\mu} + t^{-1/4} \sigma  \mathcal N& \mbox{ if } 
        \angle (\mathsf d, \mathsf u) <\theta_{\rm c}\\
        t^{\beta} X & \mbox{ if } \angle (\mathsf d, \mathsf u) >\theta_{\rm c} \,, 
    \end{cases}
\end{equation}
where $\xi_{\mu}$ is a random variable with the distribution of a constant multiple of $\mathcal Z$, hence with a heavy tail exponent $\mu$, see \eqref{heavytail}. In particular, in the typical direction, that is when the angle $\theta= \angle(\mathsf u, \mathsf d) = 0$, we have $\mu=+\infty$ and  $\xi_{\infty} \sim \mathrm{Gamma}(4\alpha)$. 
This random variable $\xi_{\mu}$ originates from  the stationary field associated with random walks in a random environment, which are tilted so as to escape ballistically as $t\mathsf u$. The standard deviation $\sigma$ of the Gaussian variable $\mathcal N$ depends on the angle and should vanish at the diagonal $\mathsf u=\mathsf d$.  
Indeed, the scale of the Gaussian fluctuations should be the same as for the functional $\mathbf I(\mathsf u, t)$ which obeys the limit theorem \eqref{eq:convergenceGaussianDrillickParekh} with the variance given by \eqref{2regimes}. The fluctuations of $\log \xi_{\mu}$, however, are not visible in the fluctuations of the quantity  $\mathbf I(\mathsf u, t)$, defined in \eqref{eq:defI},  due to the averaging  there over starting and ending points of the walk. The random variable $X$ has some probability distribution whose skewness is expected to be positive (for the KPZ class it is positive in $d=1,2$ with
$\gamma_1=0.224$ in $d=1$ and 
$\gamma_1 = 0.328$ in $d=2$ (from the 3D pt-pt DPRM/SHE in 
\cite{halpin2013extremal}). Also,
in \cite{odor2010directed} it is claimed that $\beta_{3+1} = 0.184$, which is consistent with earlier references cited therein.

Hence, the third cumulant of $\log P(t\mathsf u)$ should stay roughly constant for small angles, but it should increase as $t^{3\beta}$ for large enough angles.

We now test numericallly some of these predictions, mainly the second moment and the power law.

\subsection{Numerical results}
\subsubsection{Second moment and distribution of $\mathcal Z_{\infty}$}
\begin{figure}
\includegraphics[width=0.4\textwidth]{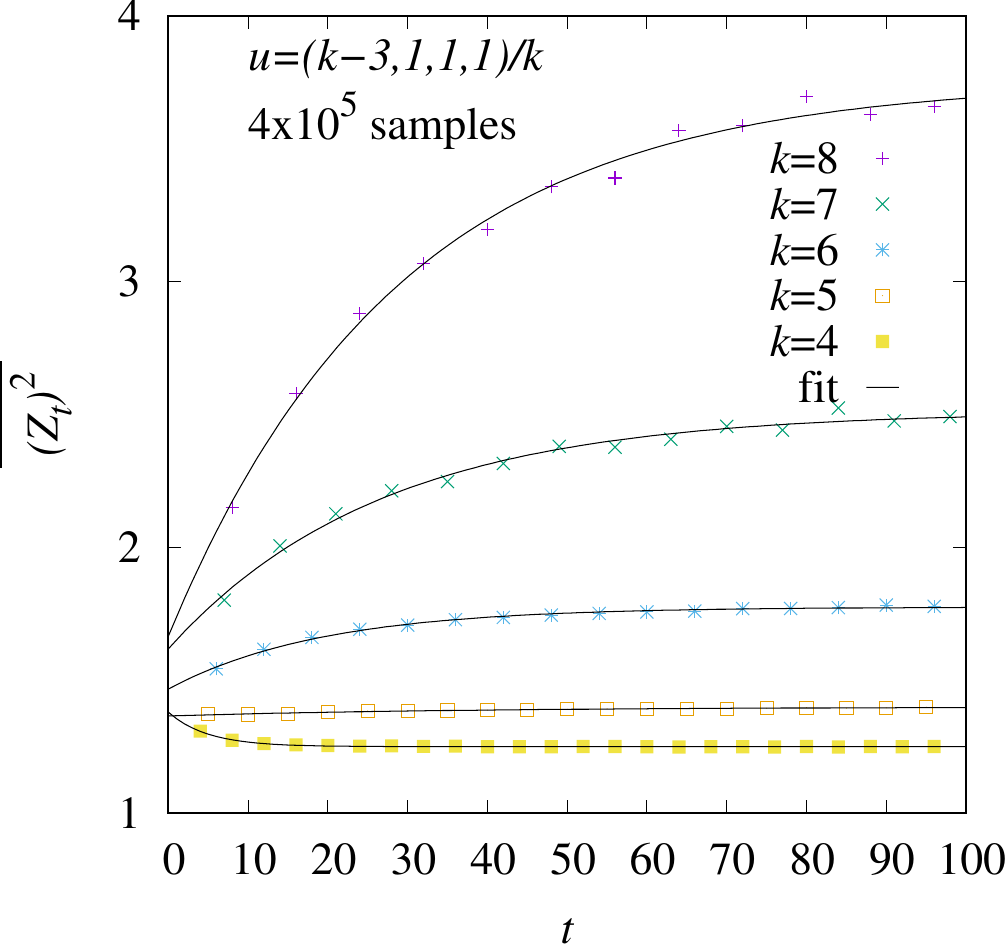}
\caption{
Second moment $\overline{(\mathcal Z_t)^2}$ as a function of time $t$, shown for values of $k=4, \dots, 8$.}
    \label{fig:second moment}
\end{figure}

We have sampled the probabilities $P(\mathsf u t)$ for lattice vectors $\mathsf u(k) =((k-3),1,1,1)/k$ and  maximum time $t_{max}=100$, for $5 \times 10^5$ time series.  The second moment $\overline{(\mathcal Z_t)^2}$ is shown in Fig. \ref{fig:second moment}. It increases towards an apparent limiting value which we estimated using a fit of the data following the form
\begin{equation}
\overline{(\mathcal Z_t)^2}=m_2(k)+b \ e^{-ct}.
\label{eq:exp:fit}
\end{equation}

\begin{table}
\begin{center} 
\begin{tabular}{|c|c|c|c|c|c|c|c|}
\hline
    $k$  &  4&5&6&7
    \\
    \hline
Simulation &     
1.2513(2) & 1.401(2) & 1.776(3) & 2.51(2)
\\ 
    \hline 
Theory    & 1.25& 1.4051& 1.82882& 2.67118
\\
\hline
\end{tabular}
\vspace*{3mm}
\begin{tabular}{|c|c|c|c|c|c|c|c|}
\hline
    $k$  &   
    8 & 9 & 10 & 11 & 11.09...\\
    \hline
Simulation &     
3.75(4) & 7.4(7) & 11(1) & 23(6) & - \\ 
    \hline 
Theory    & 
4.49578& 9.53944& 34.2218 & 4639.45 & $+ \infty$ \\
\hline
\end{tabular}
\end{center} 
\caption{\label{tab:2nd:moment}
Limiting values for second moment 
$\overline{(\mathcal Z_t)^2}$ measured along directions $\mathsf u(k)=(k-3,1,1,1)/k$ from fit to \eqref{eq:exp:fit} (second line) and compared to the analytical predictions (third line).}
\end{table}

The numerical values $m_2(k)$ obtained from the fit are given in Table \ref{tab:2nd:moment}. 
For small values of $k$, our simulation results above are very close to theoretical values.  For higher $k$, the tails of the power-law distribution \eqref{heavytail} becomes more important as the exponent $\mu=\mu(k)$ decreases. Thus, for a reliable direct estimate of the second moment and recovering the theoretical values, it would be necessary to sample substantially the tail, i.e. increase the statistics considerably.  Also one would likely have to increase the final time to obtain more precise results. Since the numerical effort grows as $O(t_{\max}^4)$ for $d=3$, this would require a too big numerical effort.

We have also tested this power law behaviour numerically, see Figure \ref{fig:powerlaw}. We see that the apparent exponent $\mu$ decreases as $k$ increases as predicted, and is just above  $2$ for $k=11$ and below $2$ for $k=12$, which signifies the divergence of the second moment just above $k=11$, which is compatible { with our exact result $k_2=11.1$.
 We also see that the measured value at $k=12$ falls not far from the one expected/conjectured at criticality $\mu=5/3=1.66$.

\begin{figure}
\includegraphics[width=0.4\textwidth]{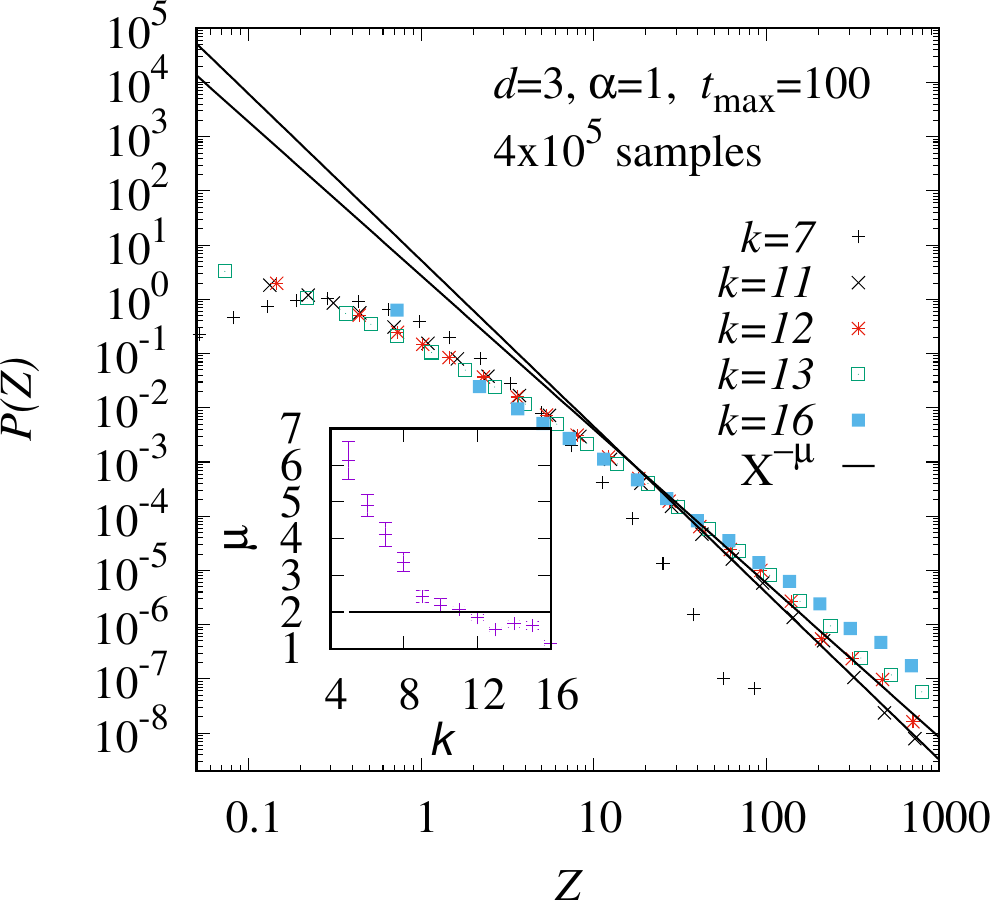}
    \caption{
    Empirical densities $\mathcal P(\mathcal Z)$ for various values of $k$. On the figure, we show the data points only for $k=7, 11, 12, 13, 16$. 
    The lines show fit of power laws $1/{\cal Z}^{1+\mu}$ to the tails for $k=11$ ($\mu=2.1(1)$)
    and $k=12$ ($\mu=1.8(1)$), respectively. 
    The inset shows all estimated values of $\mu=\mu(k)$. 
    }
    \label{fig:powerlaw}
\end{figure}

\subsubsection{Third cumulant and skewness}
We now consider the third cumulant and the skewness of $\log P(\mathsf n)$ as defined in Eqs.~(\ref{eq:third:cumulant})
and (\ref{eq:skewness}), both shown in Fig.~\ref{fig:skewLogP_k_d3_a10}.  We have considered the case  $d=3$, $\alpha=1$ and maximum time $t_{\max}=300$. We start at the diagonal $\mathsf u = \mathsf d $, which corresponds to $k=4$.

In the typical direction $k=4$, the third cumulant and the skewness are predicted to converge to 
\be \label{predictionskewness}
\kappa_3= \psi''(4 \alpha),  \quad \gamma_1= \psi''(4 \alpha)/\psi'(4 \alpha)^{3/2},
\ee 
due to the local convergence to the stationary measure. 
For $\alpha=1$ it gives $\kappa_3=-0.080$ and 
$\gamma_1=-0.52934$,
which is in pretty good agreement with the data along the diagonal: We performed a fit of a constant to the data in the range $t \in [100,300]$, i.e.,to 200 cumulant data points, each obtained from $3.7\times 10^5$ samples. This resulted in -0.5294(9), where the error bar is the statistical error bar from the fit.

\begin{figure}
\begin{center}
\includegraphics[width=0.4\textwidth]{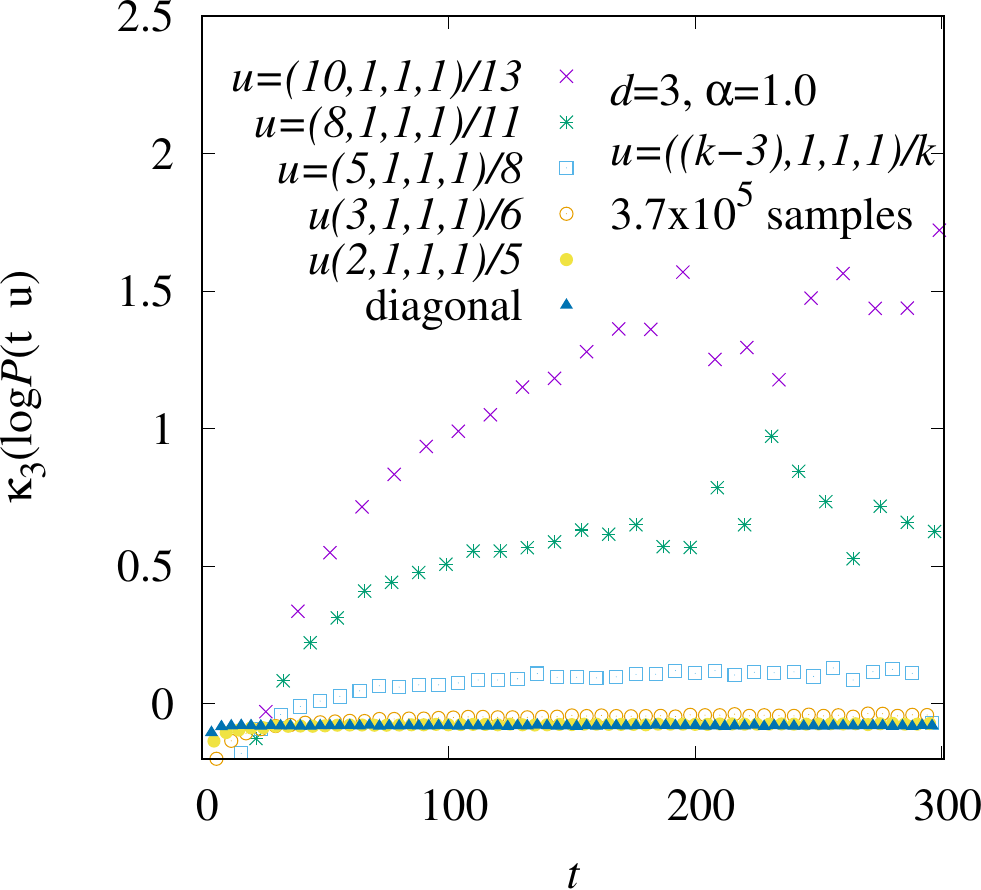}
\includegraphics[width=0.4\textwidth]{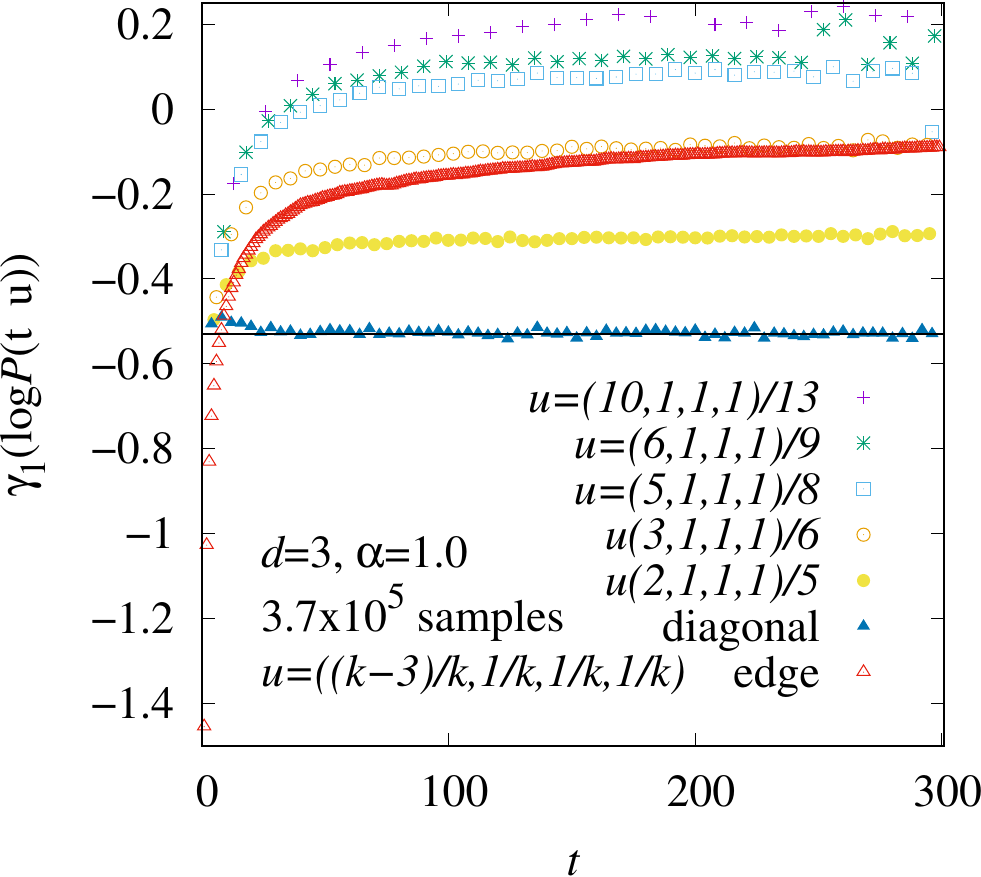}
\end{center}
\caption{\label{fig:skewLogP_k_d3_a10}
Top:  
The third cumulant of $\log P(t \mathsf u)$ for $d=3$,  $\alpha =1$, for different directions $\mathsf u$.
Bottom:  The skewness of the sample distribution of $\log P(t \mathsf u)$, for $d=3$, $\alpha =1$,
 for different directions $\mathsf u$. The horizontal line is compatible with the prediction 
 \eqref{predictionskewness}.
Note that the directions selected for the two plots differ slightly, for clarity of presentation.
}
\end{figure}

 By increasing $k$, $\mathsf u$ is moving gradually away from the diagonal as 
$\mathsf u = (2/5,1/5,1/5,1/5)$,
$\mathsf u = (3/6,1/6,1/6,1/6)$
until $\mathsf u = (10/13,1/13,1/13,1/13)$.
%
%
%
The result for $\mathsf u=(8,1,1,1)/11$ is compatible with the predictions in the weak disorder phase. These results allow to refine the understanding of the transition already observed in the variance of $\log P(\mathsf u t)$ in Section \ref{sec:variance}.

We also considered the skewness. Along the diagonal, it quickly converges to a value about $-0.5$ as predicted in \eqref{predictionskewness}.
When moving away from the diagonal, the skewness increases and becomes positive for about $\mathsf u=(4,1,1,1)/7$. For directions even farther from the diagonal, but not along an edge,
the skewness is slightly positive but does not change much. Along the edge, the behavior is different again, the
skewness is slightly negative, but correctly seems to converge to zero. This is confirmed by a plot (not shown here) of $-\gamma_1$ with log-log scale, showing a straight line of slope $-1/2$, corresponding to a power law $t^{-1/2}$.

\section{Sample to sample variance of the thermal average}

Another interesting observable is the average spatial position 
$\langle \mathsf X(t) \rangle$ in a given environment.
It is known to have fluctuations which scale non trivially
at large time, for instance $t^{1/4}$ in $d=1$ \cite{saul1992directed}. 
Here we study this observable in any dimension $d$ for the Dirichlet RWRE. 
As we argue below, it is also related to the extreme diffusion 
coefficient introduced recently in \cite{hass2024extreme}. 

\subsection{Analytical predictions}
The Dirichlet RWRE is a convenient model to compute the variance of the sample to sample fluctuations of the average deviation $\langle \mathsf X(t) \rangle$ of the random walker with respect to the diagonal. Recall that for a given $t$, it is defined as 
\begin{equation}
    \langle \mathsf X(t) \rangle 
    =\sum_{n_1+\dots + n_{d+1}=t} \mathsf n P(\mathsf n) - t\mathsf d 
\end{equation}
As $t$ goes to infinity, we have  
\be \label{eq:expectation_x_squared}
\overline{ \langle \mathsf X(t) \rangle^2 } \simeq \begin{cases}
 \frac{1}{2\alpha} \sqrt{\frac{t}{\pi}}   &\mbox{ if } d=1,\\ 
       \frac{\log t}{2\sqrt{3}\pi \alpha} &\mbox{ if } d=2,\\  
        \frac{c}{\alpha}  &\mbox{ if } d=3,
\end{cases}\ee 
where the constant $c\approx 0.336165$.  These estimates are based on 
\begin{align}
\overline{\langle \mathsf X(t)\rangle^2} &= \overline{\langle \mathsf X_1(t).\mathsf X_2(t)\rangle}\\
&= D_{\rm ext}\overline{\left\langle \sum_{s=1}^t   \mathds{1}_{\mathsf X_1(s)=\mathsf X_2(s)} \right\rangle} \label{eq:coincidences}
\end{align}
where $\mathsf X_1$ and $\mathsf X_2$ are two independent random walks in the same environment starting from the origin, and 
\be D_{\rm ext}=\overline{\langle \mathsf X_1(1)\mathsf X_2(1)\rangle} = \overline{\langle\mathsf X(1) \rangle^2}.\label{eq:Dext}\ee 
We refer to Appendix \ref{sec:appendixvariance} for the details of the derivation. Similar results are shown in \cite{saul1992directed, friedberg1994directed} for $\alpha=1$
except that Ref.~\cite{friedberg1994directed} considers a
slightly different model (where at each step, the random walk $\mathsf X(t)$ makes a step among the $2d$ directions $\pm \mathsf e_1, \dots, \pm \mathsf e_d$).

\subsection{Numerical results}

In Fig.~\ref{fig:Ex2_d1} the numerical result of $\overline{ \langle \mathsf X(t) \rangle^2}$
for the case $d=1$ is shown for the four values $\alpha=0.2,0.5, 1$, and 2. An average over 10000 samples of the environment was performed. The rescaling of the $y$-axis by $\alpha$ leads to a collapse of the data for large times $t$, confirming
the $1/\alpha$ dependence of the prefactor as given by Eq.~(\ref{eq:expectation_x_squared}). Also,
for comparison, the predicted function is shown, which confirms the long-time behavior. 

\begin{figure}
\begin{center}
\includegraphics[width=0.4\textwidth]{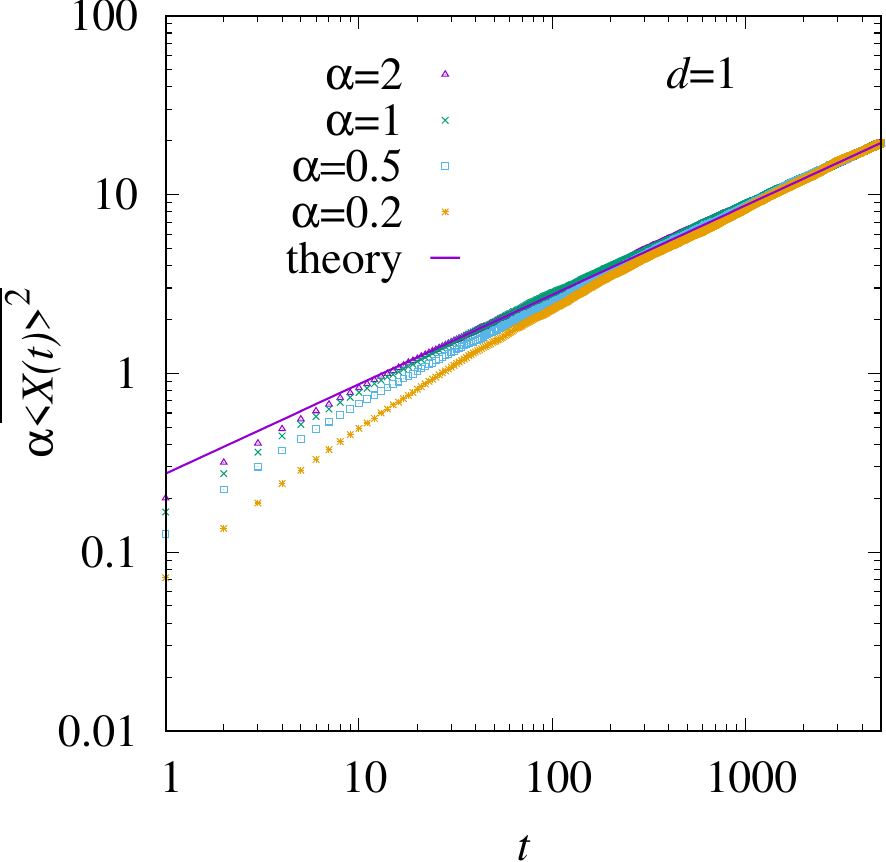}
\end{center}
\caption{\label{fig:Ex2_d1}
The ensemble expectation of the squared average walk position multiplied
by $\alpha$, i.e., 
$\alpha \overline{ \langle \mathsf X(t) \rangle^2}$,
as a function
of the time $t$ for $d=1$. The line shows the analytical result Eq.~(\ref{eq:expectation_x_squared}) for comparison.
}
\end{figure}

The corresponding result for $d=2$ is shown in Fig.~\ref{fig:Ex2_d2}. The $\log(t)$
scaling of the time axis leads to a linear behavior, which confirms the $\log(t)$
scaling of Eq.~(\ref{eq:expectation_x_squared}). Here
also an $1/\alpha$ behavior can be observed, although the fluctuations
are a bit large for large values of $t$ making the collapse look less perfect. In particular 
if one omits (not shown) the  case $\alpha=0.2$, the collapse looks somehow cleaner.
The predicted curve  Eq.~(\ref{eq:expectation_x_squared}) matches well if it is shifted by about 0.25 up, which corresponds to rescaling the time axis. For the plot we have shifted by 0.27 for better visibility. Thus, the predicted square root behavior, the $1/\alpha$
scaling and the other constants in front of the log match the data very well.

\begin{figure}
\begin{center}
\includegraphics[width=0.4\textwidth]{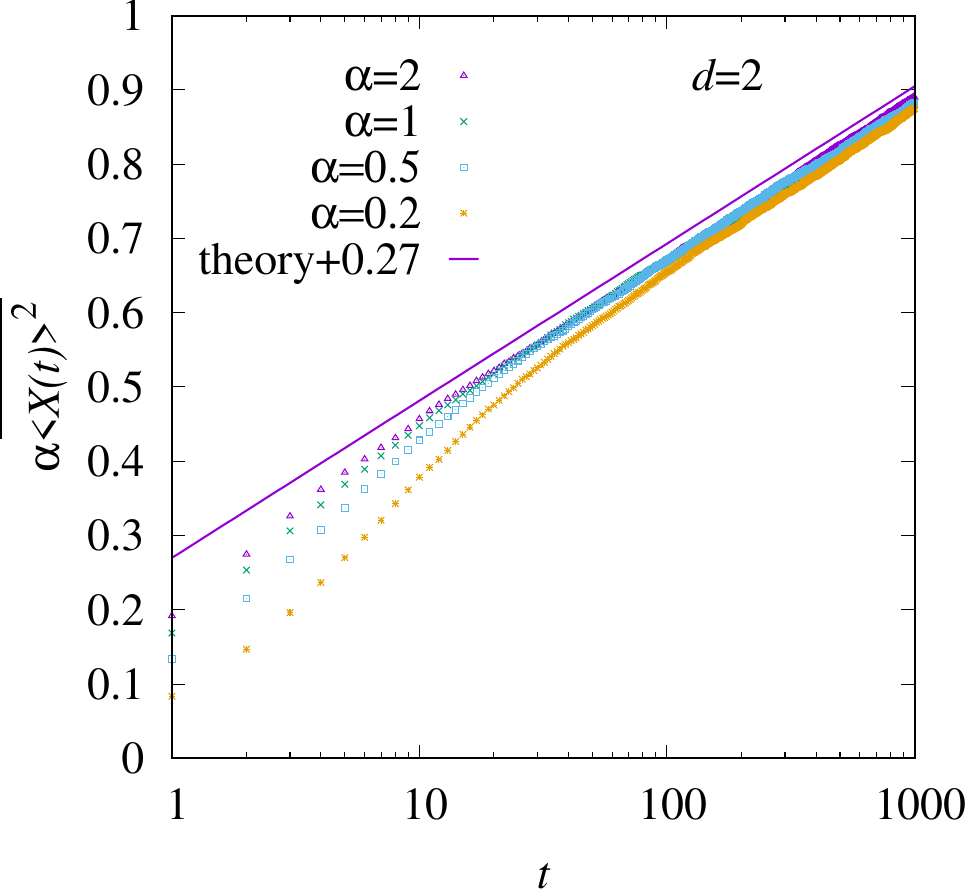}
\end{center}
\caption{\label{fig:Ex2_d2}
The ensemble expectation of the squared average walk position multiplied
by $\alpha$, i.e., 
$\alpha \overline{ \langle \mathsf X(t) \rangle^2}$,
as a function
of the time $t$ for $d=2$ with logarithmically scaled $t$ axis.
The line shows the expected behavior Eq.~(\ref{eq:expectation_x_squared}) but shifted
up by 0.27, for comparison.}
\end{figure}

In Fig.~\ref{fig:Ex2_d3} the result for $d=3$ is shown. With increasing
time $t$, a convergence towards a constant is compatible with the data, as stated in 
Eq.~(\ref{eq:expectation_x_squared}). Still, due to the large dimension, we
cannot reach large enough times $t$ to actually observe the
convergence. To verify this, we have fit the data of $\alpha=1$ to a power
law plus constant, i.e.,
\begin{equation}
\label{eq:power:fit}
    f(t)=c+b t^{-a}\,.
\end{equation}
This resulted in a very good
fit, see inset of the figure, for all times $t\ge 1$, with $c=0.3351(1), b=-0.183(1)$ and $a=0.613(4)$. The fitted asymptotic value of $c$ is 
close to the predicted value $0.336165$.
For better visibility, we show $f(t)+0.01$ in the main plot.
Here, for $\alpha \ge 0.5$ we again
observe a good collapse of the data, if rescaled by $\alpha$. For the
small value $\alpha=0.2$, the rescaled data is a bit off.
\begin{figure}
\begin{center}
\includegraphics[width=0.4\textwidth]{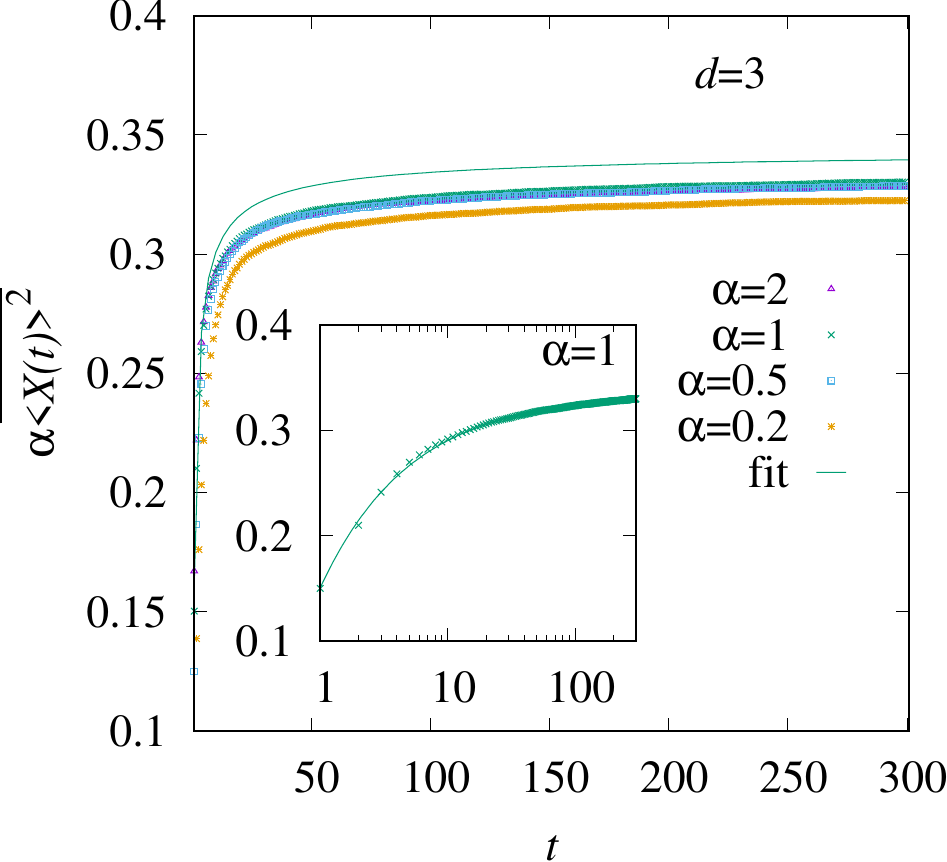}
\end{center}
\caption{\label{fig:Ex2_d3}
The ensemble expectation of the squared average walk position 
multiplied
by $\alpha$, i.e., 
$\alpha \overline{ \langle \mathsf X(t) \rangle^2}$,
as a function
of the time $t$ for $d=3$. A convergence towards a constant is compatible with the data.
The line shows a fit to a power law Eq.~(\ref{eq:power:fit}), shifted by 0.01 for
better visibility. The inset shows the data for $\alpha=1$ and the same fit,
not shifted, with
a logarithmically scaled $t$ axis, indicating the very good agreement of the fit with the data.}
\end{figure}

\subsection{Connection to the extreme diffusion coefficient}
An extreme diffusion coefficient 
\be 
\lambda_{\rm ext} = \frac{1}{2}\frac{D_{\rm ext}}{D-D_{\rm ext}},
\ee 
where $D_{\rm ext}$ is defined in \eqref{eq:Dext},  
is introduced in \cite{hass2024extreme} for a class of models of RWRE in $d=1$,
(to which the present model belongs \footnote{In references \cite{hass2024extreme, hass2025universal}, the diffusion coefficients are defined with an extra factor $1/2$ which has no effect on the definition of $\lambda_{\rm ext}$.}). It takes into account the global effect of the local random drift imposed to random walkers in random environment.  For more general RWRE, e.g. when the environment is correlated in space,  the coefficient has a more complicated expression given in \cite[Eq. (5)]{hass2025universal} (see also \cite{parekh2024hierarchy}). Comparing \cite[Eq (22)]{hass2025universal} with \eqref{eq:coincidences}, we note that when the environment is uncorrelated in space, the extreme diffusion coefficient is in fact related to the sample to sample variance of the thermal average. We have  
\be 
\lim_{t\to\infty} \frac{\overline{ \langle \mathsf X(t) \rangle^2}}{\sqrt{D t}} 
= \lambda_{\rm ext} \overline{L_0^{B_1-B_2}(1)}  = \lambda_{\rm ext}  \frac{1}{\sqrt{\pi}}
\ee 
where $L_0^{B_1-B_2}(1)$ denotes the local time at $0$ of the difference of two standard Brownian motions, up to time $1$. 
The same coefficient $\lambda_{\rm ext}$ is denoted by $\gamma_{\rm ext}^2$ in \cite{parekh2024hierarchy}.  It is shown in \cite{drillick2025random} that the same coefficient governs the variance of the noise both in the weak disorder regime and in the moderate deviation regime.

Finally, let us indicate a heuristic argument which, in any dimension,  relates
the moderate deviation/critical scale to the observable
$\overline{ \langle \mathsf X(t) \rangle^2}$. 
In a given environment, we approximate 
\be 
\mathbf P(\mathsf X(t)=\mathsf x) \approx  \frac{1}{(2\pi \tilde D t)^{d/2}} e^{- \frac{(\mathsf x - \langle \mathsf X(t) \rangle)^2}{2 \tilde D t} }.
\ee 
Expanding inside the exponential we see that the free diffusion is biased
by the measure 
\be 
e^{ \frac{1}{\tilde D t} \mathsf x\langle \mathsf X(t) \rangle } \,.
\ee 
While it is negligible for $\vert \mathsf x\vert  \asymp \sqrt{t}$ it becomes 
of order $1$ on scales 
\be 
\vert \mathsf x \vert \asymp t/ \sqrt{\mathrm{Var} \langle \mathsf X(t) \rangle}
\label{eq:criticalscale}
\ee
which coincides with the moderate deviation/critical scale.
Hence at that scale fluctuations of $\log \mathbf P(\mathsf X(t)=\mathsf x)$ due to the environment become of order $1$.  A more precise argument is provided in \cite{drillick2025random}, relating the critical scale with the local time of intersection of two RWRE $\sum_{s=1}^t\mathds{1}_{X_1(s)=X_2(s)}$. For the Dirichlet RWRE,  this local time is related by \eqref{eq:coincidences} to the variance of $\langle \mathsf X(t) \rangle$ so that we recover \eqref{eq:criticalscale}.

\section{Summary and Discussion}

We have studied numerically the behavior of a RWRE with Dirichlet-distributed steps
in $d+1$ dimensions for $d=1,2$ and 3, with uniform Dirichlet parameter
$\alpha$. Since we store only the position distributions for two consecutive
times, large final times can be considered, depending on the dimension $d$.
In particular we measure
the disorder-driven distributions of the probability $P(\mathsf n)$ to reach
at time $t$ certain positions $\mathsf R(t)=\mathsf n$ along lines $\mathsf n=\mathsf u t$. Here, $\mathsf u$ are
vectors $\mathsf u=(u_1,\ldots,u_{d+1})$ with $\sum_i  u_i = 1$ in directions
along the diagonal, slightly and more stronger off diagonal, and along the
edges of the $d+1$ sub space of accessible sites. 

 The advantage of the Dirichlet RWRE is that we could determine
its exact stationary measure, which allows to make predictions
about the distribution of $\log P(\mathsf n)$ along the typical
direction, i.e the diagonal. Indeed, along the diagonal,
we find that for all dimensions the variance and third cumulant of $\log P(\mathsf n)$
converges to constants, with a very good agreement to the theoretical
calculations. Off diagonal, for $d=1,2$, we observe a growing behavior
$\log P(\mathsf n) \sim t ^{2 \beta}$, where $\beta$ is the KPZ roughness
exponent in $d+1$ dimensions, also as analytically predicted.
For $d=3$ this is only happening beyond a certain critical
angle, i.e in the strong disorder phase. 
More precisely, we have considered the vectors 
$\mathsf u= (k-3,1,1,1)/k$. For $k\leq 11$ we found that
the variance and third cumulant of $\log P(\mathsf ut)$ saturates at 
large time, a signature of the weak disorder phase. 
For $k=13$ we obtain evidence that these cumulant grow with time,
as predicted in the strong disorder phase. Note that
the largest time $t_{\max}$ accessible to us is not sufficient to probe further the precise
value of the exponent $\beta$, as well as the asymptotic skewness
and the precise value of $k_c$. 

In addition, we obtained a more detailed picture of the weak disorder phase by studying the observable
$\mathcal{Z}_t= P(\mathsf u t)/\overline{P(\mathsf u t)}$. 
We obtained
an exact formula for 
$\overline{\mathcal{Z}_{\infty}^2}$
which provides a lower bound on the transition $k_c \geq 11.1$.
Using recent mathematical results on directed polymers, we also argued and showed numerically that the random
variable 
$\mathcal{Z}_{\infty}$ acquires a heavy tail distribution.

Finally, we have studied the sample to sample variance of
the mean position, 
$\overline{ \langle \mathsf X(t) \rangle^2 }$.
The numerics is in good agreement with our analytic
calculation of its large time asymptotics. We have
related this observable to the extremal diffusion
coefficient (within the present model) and
argued that it allows to predict the correct critical/moderate
deviation scale in $d=1,2,3$.


\begin{acknowledgments}
We thank Shalin Parekh for a useful discussion related to the results of \cite{drillick2025random} and the stationary measures of random walks in random environment. We also thank Ivan Corwin, Hindy Drillick and Jacob Hass for discussions about RWRE in $d \geq 1$. 
  The simulations were performed at the
  the HPC cluster ROSA, located at the University of Oldenburg
  (Germany) and
    funded by the DFG through its Major Research Instrumentation Program
    (INST 184/225-1 FUGG) and the Ministry of
    Science and Culture (MWK) of the
    Lower Saxony State. G. B acknowledges support from ANR grants ANR-21-CE40-0019 and
    ANR-23-ERCB-0007. 
     PLD acknowledges support from ANR Grant No. ANR23-CE30-0020-01 EDIPS.
     This research was supported in part by grant NSF PHY-2309135 to the Kavli Institute for Theoretical Physics (KITP). 
\end{acknowledgments}

\vspace*{3mm}

\centerline{\bf Data availability statement}

\vspace*{3mm}

The data that support the findings of this article are openly available
\cite{data_sources2026}.

\bibliography{refs.bib}

\newpage 
\appendix
\begin{widetext}
\section{Sample to sample variance of the thermal average}
\label{sec:thermalaverage}
In this section, we explain how to obtain the estimate \eqref{eq:expectation_x_squared} for the sample to sample variance of the thermal average. 
\label{sec:appendixvariance}
\subsection{Dimension $1$} 
Recall that $\mathsf d=(1/2,1/2)$ denotes the diagonal direction. Consider two random walks $ \mathsf X_1(t), \mathsf X_2(t)$ in the same environment. 
The walks make steps in the directions 
$$ \mathsf s_1 =\frac{1}{2}(1,-1) , \;\;\mathsf s_2 =\frac{1}{2}(-1,1).$$
Since $\mathsf s_1 \cdot\mathsf s_1=\frac{1}{2}$ and $\mathsf s_1 \cdot\mathsf s_2=\frac{-1}{2}$, we have 
\be 
\overline{ \left\langle \mathsf X_1(t)\cdot \mathsf X_2(t)\right\rangle} =  \sum_{s=0}^{t-1} \overline{\mathbf P(\mathsf X_1(s)= \mathsf X_2(s))} \times  \overline{\tfrac{1}{2} (2 p-1)^2 },\quad \quad \overline{ (2 p-1)^2 }  = \frac{1}{2 \alpha +1} 
\ee 
where $\frac{1}{2}(2 p-1)^2 = \frac{1}{2}\left(p^2 + (1-p)^2\right) - \frac{1}{2} 2 p(1-p)$ is the covariance of the increments $\mathsf X_1(s+1)-\mathsf X_1(s)$ and $\mathsf X_2(s+1)- \mathsf X_2(s)$ at any time $s$, whenever the paths start from the same point, i.e., $\mathsf X_1(s)=\mathsf X_2(s)$.

In order to compute 
\begin{equation} 
\sum_{s=0}^{t-1} \overline{\mathbf P(\mathsf X_1(s)=\mathsf X_2(s))} = \overline{\left\langle\sum_{s=0}^{t-1} \mathds{1}_{\mathsf X_1(s)=\mathsf X_2(s)} \right\rangle}, 
\end{equation}
we consider an auxiliary random walk $Y$ with the same transition probabilities as $\mathsf X_1-\mathsf X_2$, after averaging over the environment. This walk makes steps $(0,0)$, $(1,-1)$, and $(-1,1)$, i.e., it moves with steps $0, +1, -1$ in a fixed direction. If $Y\neq 0$, these steps occur with probabilities $1/2, 1/4, 1/4$ and when $Y=0$, the same steps occur with probability $\overline {p^2+(1-p)^2}= \frac{1+\alpha}{1 + 2 \alpha}$, $\overline{p(1-p)}= \frac{\alpha}{2(1 + 2 \alpha)}$ and 
$\overline{p(1-p)}$, respectively. It will be convenient to let $q$ be the probability to stay put, i.e.,
\begin{equation}
q:= \overline{p^2+(1-p)^2}= \frac{1+\alpha}{1 + 2 \alpha} = 1-\frac{\alpha}{2\alpha+1}.
\end{equation}
Let us decompose the trajectory of a sample walk $Y$. Initially, the walk stays at zero for some time $1+G_1$, where $G_1\sim\mathrm{Geom}(q)$, a geometric random variable with parameter $q$.  We recall the definition $P(G_1=n)= (1-q) q^n$,$n \geq 0$.

Then the walk does an excursion staying above $1$ or below $-1$ for some time $E_1$, until a random time $T_1$. Then again, it will stay for some time $1+G_2$ at zero, then make an excursion of length $E_2$, etc. 

For any $t$, we may consider the random number $N$ such that $T_{N-1}< t\leq T_N$, i.e.,that the time $t$ is in the $N$-th excursion. The  amount of time the walk $Y$ spends at zero is 
\begin{equation}
\sum_{s=0}^{t-1} \mathds{1}_{Y(s)=0} =    \sum_{i=1}^{N} (1+G_i)   \simeq\, N\, \frac{2\alpha+1}{\alpha}.
\end{equation}
 Hence, in principle we need to estimate the asymptotics of $N=N(t)$ when $t$ is large. Instead of studying directly this quantity, it is useful to compare the walk $Y$ with the lazy random walk $\overline Y$ which jumps to  $+1, 0$ or $-1$ with probabilities $1/4,1/2,1/4$. The walks $Y$ and $\overline Y$ can be coupled together in such a way that both walks make the same excursions away from zero, and the only difference being between two excursions: $Y$ stays at zero for a time distributed as $1+\mathrm{Geom}(q)$ while $\overline Y$ stays at zero for a time distributed as $1+\mathrm{Geom}(1/2)$. After a large number of excursions, the total  time spend at zero is proportional to the average of those geometric random variables.  We conclude that, as $t$ goes to infinity,  
\begin{equation}
 \overline{\left\langle\sum_{s=0}^{t-1} \mathds{1}_{\mathsf X_1(s)=\mathsf X_2(s)} \right\rangle}   \simeq \frac{\overline{1+\mathrm{Geom}(q)}}{\overline{1+\mathrm{Geom}(1/2)}} \sum_{s=1}^t P_s,
\end{equation}
where $P_s$ is the probability to be at zero for the lazy random walk $\overline Y$. This approximation is correct as long as  $N(t)=\overline N(t)+ o(N(t))$ as $t$ goes to infinity, where $\overline N$ is the number of excursions made by the walk $\overline Y$. 
 We have 
\begin{equation}
    P_s = \frac{1}{2\I\pi}\oint \left( \frac{z+z^{-1}+2}{4} \right)^s \frac{dz}{z}. 
\end{equation}
Summing over $s$ gives  
\begin{equation}
 \sum_{s=0}^{t-1} P_s =  \frac{-4}{2\I\pi}\oint \frac{1-e^{t\log\left(\frac{(z+1)(z^{-1}+1}{4} \right)}}{(z-1)^2}dz
 \end{equation}
 and applying  Laplace's method (i.e.,apply the change of variables $z=1+\frac{1}{\sqrt{t}}\I x$, $x\in \R$), we find 
\begin{equation}
    \sum_{s=1}^t P_s \simeq \frac{4 \sqrt{t}}{2 \pi} \int_{-\infty}^{+\infty}  dx \frac{1- e^{-x^2/4}}{x^2} 
    =
    \frac{2\sqrt{t}}{\sqrt{\pi}}.
\end{equation}
Thus, we finally arrive at 
\begin{equation}
\boxed{   \overline{ \left\langle \mathsf X_1(t). \mathsf X_2(t)\right\rangle} \sim \frac{\sqrt{t}}{2\alpha\sqrt{\pi}}.}
\end{equation}
\begin{remark} For a general symmetric one-dimensional random walk,  one has 
\begin{equation}
    P_s = \frac{1}{2\I\pi}\oint \left( \sum_{k \geq 1}  
    p_k (z^k + z^{-k}) + 1 - 2 \sum_{k \geq 1}  p_k \right)^s \frac{dz}{z}
\end{equation}
Following the same steps one finds 
\be 
P_s \simeq \frac{1}{\sqrt{s}} \int_{-\infty}^{+\infty}  
\frac{dx}{2 \pi} e^{- x^2 \sum_{k \geq 1}  k^2 p_k } = 
\frac{1}{2 \sqrt{\sum_{k \geq 1}  k^2 p_k }} 
\frac{1}{ \sqrt{ \pi s}} = \frac{1}{\sqrt{2 \pi D s} },  \quad  \quad D= 2 \sum_{k \geq 1}  k^2 p_k 
\ee 
\end{remark}

\subsection{Dimension $2$}
In dimension $d=2$, walks make steps $\mathsf e_1, \mathsf e_2,\mathsf e_3 $ in $\Z^{d+1}$. In the plane $\mathbb T_2$ orthogonal to the diagonal direction $\mathsf d=(1/3,1/3,1/3)$, the random walk  steps are 
\begin{equation} 
\mathsf s_1= \mathsf e_1-\mathsf d=\frac{1}{3}(2,-1,-1), \; \mathsf s_2= \mathsf e_2-\mathsf d=\frac{1}{3}(-1, 2,-1), \; \mathsf s_3= \mathsf e_3-\mathsf d=\frac{1}{3}(-1,-1, 2).
\end{equation}
We have $\mathsf s_i \cdot \mathsf s_i=2/3$ and $\mathsf s_i \cdot \mathsf s_i=-1/3$ for $i\neq j$. Decomposing the trajectory as a sum of increments, we obtain 
\begin{equation}
    \overline{\left\langle  \mathsf X_1(t).\mathsf X_2(t)\right\rangle} =   \sum_{s=0}^{t-1} \overline{\mathbf P(\mathsf X_1(s)=\mathsf X_2(s))}\times  \overline{ \frac{2}{3}(p_1^2+p_2^2+p_3^2) -\frac{1}{3}(2p_1p_2+2p_2p_3+2p_1p_3)}, 
\end{equation}
where $p_1,p_2,p_3$ are the probabilities to make steps in directions $\mathsf e_1, \mathsf e_2, \mathsf e_3$. Regarding the time spent at the origin for the walk $Y$ (defined again as the random walk with the same transition probabilities  as $\mathsf X_1-\mathsf X_2$ after averaging over the environment), we will do the same sort of decomposition in excursions as when $d=1$, except that now, the parameter of the geometric variable is $q= \mathbb E[p_1^2+p_2^2+p_3^2]$ and the number of excursions is   $c\log(t)$. 
For  $(p_1,p_2,p_3)\sim \mathrm{Dir}(\alpha, \alpha, \alpha)$, we have 
\begin{equation}
\overline{p_1^2} =\overline{p_2^2} =\overline{p_3^2} = \frac{\alpha(\alpha+1)}{3\alpha(3\alpha+1)} 
\end{equation}
so that 
\begin{equation} \overline{1+\mathrm{Geom}(q)} = \frac{1}{1-q} = \frac{3\alpha+1}{2\alpha}.
\end{equation}
Furthermore, 
\begin{equation}
    \frac{2}{3}\overline{p_1^2+p_2^2+p_3^2} -\frac{1}{3} \overline{2p_1p_2+2p_2p_3+2p_1p_3} = \frac{2}{3(3\alpha+1)}
\end{equation}
so that finally, 
\begin{equation}
\overline{\left\langle  \mathsf X_1(t) \cdot \mathsf X_2(t)\right\rangle}= \frac{2}{3(3\alpha+1)}\frac{3\alpha+1}{2\alpha} \overline{N(t)} \sim \frac{c \log t}{3\alpha }
\end{equation}

 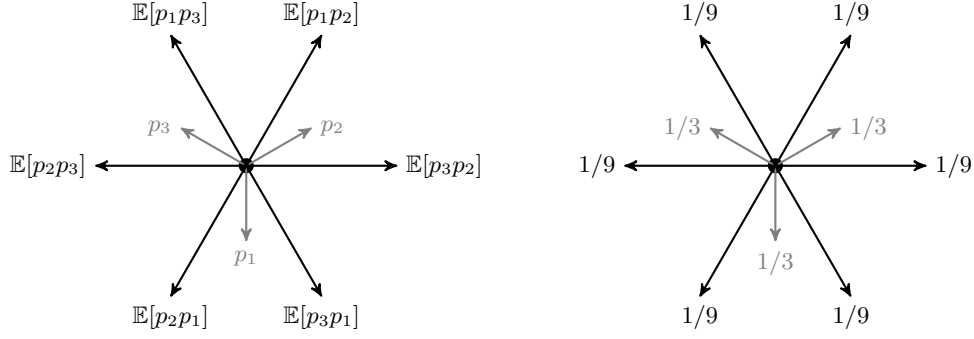
\begin{figure}
    \centering
    \begin{tikzpicture}
    \fill (0,0) circle(0.1);
    \draw[thick, gray, -stealth'] (0,0) -- (30:1) node[right]{$p_2$};
    \draw[thick, gray, -stealth'] (0,0) --({30+120}:1) node[left]{$p_3$};
    \draw[thick, gray, -stealth'] (0,0) -- ({30+240}:1) node[below]{$p_1$};
    \foreach \x in {0,1,...,5}{
    \draw[thick, -stealth'] (0,0) -- ({\x*60}:2);
    }
\draw ({0*60}:2) node[right]{$\mathbb E[p_3p_2]$};
\draw ({1*60}:2) node[above]{$\mathbb E[p_1p_2]$};
\draw ({2*60}:2) node[above]{$\mathbb E[p_1p_3]$};
\draw ({3*60}:2) node[left]{$\mathbb E[p_2p_3]$};
\draw ({4*60}:2) node[below]{$\mathbb E[p_2p_1]$};
\draw ({5*60}:2) node[below]{$\mathbb E[p_3p_1]$};
\begin{scope}[xshift=7cm]
        \fill (0,0) circle(0.1);
    \draw[thick, gray, -stealth'] (0,0) -- (30:1) node[right]{$1/3$};
    \draw[thick, gray, -stealth'] (0,0) --({30+120}:1) node[left]{$1/3$};
    \draw[thick, gray, -stealth'] (0,0) -- ({30+240}:1) node[below]{$1/3$};
    \foreach \x in {0,1,...,5}{
    \draw[thick, -stealth'] (0,0) -- ({\x*60}:2);
    }
\draw ({0*60}:2) node[right]{$1/9$};
\draw ({1*60}:2) node[above]{$1/9$};
\draw ({2*60}:2) node[above]{$1/9$};
\draw ({3*60}:2) node[left]{$1/9$};
\draw ({4*60}:2) node[below]{$1/9$};
\draw ({5*60}:2) node[below]{$1/9$};
\end{scope}
    \end{tikzpicture}
    \caption{Each walk $\mathsf X_1$ and $\mathsf X_2$ can move in three directions (in gray) at each step. As a result, the walk $Y=\mathsf X_1-\mathsf X_2$ can move in the plane $\mathbb T_2$ orthogonal to $\mathsf d$ in the $6$ directions indicated with black arrows. The probabilities in the left picture correspond to the transition probabilities when $Y=0$ (the probability to stay put is $\mathbb E[p_1^2+p_2^2+p_3^2]$), while the right picture indicates transition probabilities when $Y\neq 0$ (the probability to stay put is $1/3$).}
    \label{fig:dimension2}
\end{figure}

Instead of estimating directly the number $N(t)$ of excursions of the walk $\mathsf Y=\mathsf X_1-\mathsf X_2$ after averaging over the environment, we use a similar argument as above.  We consider the modified random walk $\overline Y$ defined as  the random 
 walk on the triangular lattice which stays put with probability $1/3$ and jumps to one of the six  neighbours with probability $1/9$ (i.e.,the walk with transition probabilities depicted in Figure \ref{fig:dimension2} (right)). 
Hence we have 
\begin{equation}
\begin{split}  
   \overline{\left\langle  \mathsf X_1(t).\mathsf X_2(t)\right\rangle} &=  \frac{2}{3(3\alpha+1)} \sum_{s=0}^{t-1} \overline{\mathbf P(\mathsf X_1(s)=X_2(s))},\\
   &\simeq \frac{2}{3(3\alpha+1)} \frac{\overline{1+\mathrm{Geom}(q)}}{\overline{1+\mathrm{Geom}(1/3)}} \sum_{s=0}^{t-1} P_s,\\
   &\simeq \frac{2}{9\alpha}\sum_{s=0}^{t-1} P_s 
\end{split}
\label{eq:estimated=2}
\end{equation}
where $P_s$ is the probability for for the walk $\overline Y$ to  be at zero at time $s$. It can be computed as  
\begin{equation}
    P_s = \frac{1}{(2\I\pi)^2} \oint\frac{\mathrm dz}{z} \oint \frac{\mathrm dw}{w} \left(\frac{1}{9}\left( z+w+z/w+w/z+1/w+1/z\right) +1/3\right)^t
\end{equation}
Using the change of variables $z=1+\frac{1}{\sqrt{t}} \I x, w=1+\frac{1}{\sqrt{t}} \I y$, we obtain that as $t$ goes to infinity, 
\begin{equation}
    P_t \sim \frac{1}{(2\pi)^2 t} \int_{\mathbb R} dx \int_{\mathbb R} dy e^{\frac{-2}{9}(x^2-xy+y^2)} = \frac{3\sqrt{3}}{4\pi t}.
\end{equation}
Thus, 
\begin{equation}
     \sum_{s=0}^t P_s \sim \frac{3\sqrt{3}}{4\pi} \log(t).
     \label{eq:returns-random-walk-d=2}
\end{equation}
Finally, we conclude from \eqref{eq:estimated=2} together with  \eqref{eq:returns-random-walk-d=2} that 
\begin{equation}
   \boxed{  \overline{\left\langle  \mathsf X_1(t) . \mathsf X_2(t)\right\rangle} \simeq  \frac{\log (t)}{2\sqrt{3} \pi  \alpha}.}
\end{equation}

\subsection{Dimension $3$}
In dimension $d=3$, the random walk steps in the hyperplane $\mathbb T_3$ orthogonal to $\mathsf d$ are 
$$ \mathsf s_1=\frac{1}{4}(3,-1,-1, -1), \; \mathsf s_2=\frac{1}{4}(-1,3,-1,-1), \; \mathsf s_3=\frac{1}{4}(-1,-1, 3, -1), \; \mathsf s_4=\frac{1}{4}(-1,-1, -1, 3)$$
so that the covariance of increments when walks start from the same point is 
\begin{equation} 
    \mathsf s_1 \cdot \mathsf s_1 q + \mathsf s_1 \cdot \mathsf s_2 (1-q) = \frac{3}{4} q -\frac{1}{4}(1-q) 
\end{equation}
where $q= \overline{p_1^2+p_2^2+p_3^2+p_4^2} = \frac{\alpha+1}{4\alpha+1}$.
Hence, following the same argument as in lower dimensions,  we obtain that 
\begin{equation}
    \begin{split}
      \overline{\left\langle  \mathsf X_1(t).\mathsf X_2(t)\right\rangle}  &=  \frac{3}{4(4\alpha+1)}   \sum_{s=0}^{t-1}  \overline{\mathbf P(\mathsf X_1(s)=\mathsf X_2(s))}, \\ 
       &\simeq   \frac{3}{4(4\alpha+1)}   \frac{\overline{1+\mathrm{Geom}(q)}}{\overline{1+\mathrm{Geom}(1/4)}} \sum_{s=0}^{t-1} P_s,\\
       &\simeq \frac{3}{16\alpha} \sum_{s=0}^{t-1} P_s
    \end{split}
    \label{eq:estimated=3}
\end{equation}
where $P_s$ is the probability to be at zero at time $s$ for the walk $\overline Y$ (again,  $\overline Y$ denotes the modification of the walk $Y$ which has the same transitions at zero and away from zero). This probability may be computed as 
\begin{equation}
    P_s = \frac{1}{(2\I\pi)^3} \oint\frac{\mathrm dz_1}{z_1} \oint \frac{\mathrm dz_2}{z_2}\oint \frac{\mathrm dz_3}{z_3} \left(f(z_1,z_2,z_3)\right)^s
    \label{eq:integralreturnproba}
\end{equation}
where 
\begin{align} 
f(z_1,z_2,z_3)&:= \frac{1}{16}\left( z_1+\frac{1}{z_1}+z_2+\frac{1}{z_2}+z_3+\frac{1}{z_3}+\frac{z_1}{z_2} +\frac{z_2}{z_1} +\frac{z_1}{z_3} +\frac{z_3}{z_1} + \frac{z_2}{z_3} +\frac{z_3}{z_2} \right) +1/4 \\
&= \frac{1}{16} \left( 1+z_1+z_2+z_3\right)\left( 1 +\frac{1}{z_1} +\frac{1}{z_2} +\frac{1}{z_3}\right).\end{align}
Summing over $s$, we get 
\begin{equation}
    \sum_{s=0}^{t-1} P_s = \frac{1}{(2\I\pi)^3}\oint\frac{\mathrm dz_1}{z_1} \oint \frac{\mathrm dz_2}{z_2}\oint \frac{\mathrm dz_3}{z_3} \frac{1- f(z_1,z_2,z_3)^t}{1-f(z_1,z_2,z_3)}, 
\end{equation}
where the numerator has a singularity at $z_1=z_2=z_3=1$ which cancels with that of the numerator. It is not difficult to show that $P_t=O(t^{-3/2})$ so that $ \sum_{s=0}^t P_s$ converges to a constant as $t$ goes to infinity given by 
\begin{multline}  \frac{1}{(2\I\pi)^3} \oint\frac{\mathrm dz_1}{z_1} \oint \frac{\mathrm dz_2}{z_2}\oint \frac{\mathrm dz_3}{z_3} \frac{1}{1-f(z_1,z_2,z_3)} \\
 = 
\prod_{i=1}^3 \int_0^{2 \pi} \frac{d\theta_i}{2 \pi}    
 \frac{8}{6- \cos \left(\theta
   _1\right)-\cos \left(\theta
   _1-\theta _2\right)-\cos
   \left(\theta _2\right)-\cos
   \left(\theta _1-\theta
   _3\right)-\cos \left(\theta
   _2-\theta _3\right)-\cos
   \left(\theta _3\right)} \\
   = 1.79288
   \label{eq:L(0)}
\end{multline}
Thus, using \eqref{eq:estimated=3}, we find 
\begin{equation}
  \boxed{  \overline{\left\langle  \mathsf X_1(t) . \mathsf X_2(t)\right\rangle}  \xrightarrow[t\to\infty]{} \frac{c}{\alpha} \text{ with } c\approx   0.336165.}
\end{equation}

\section{Calculation of the second moment in dimension $d\geq 3$}
\label{sec:secondmoment}
The goal of this section is to analyze the asymptotics of the rescaled second moment 
\be \overline{ \mathcal Z_t^2} = \frac{\overline{\mathbf P(\mathsf X(t)=t v \mathsf s_1)^2}}{\left(\overline{\mathbf P(\mathsf X(t)=t v \mathsf s_1)}\right)^2}
\label{eq:rescaledsecondmoment}
\ee
for a velocity $v\in (0,1)$.
\subsection{First moment} 
Let us start with the computation of the first moment $\overline{\mathbf P(\mathsf X(t)=t v \mathsf s_1)}$. In order to arrive at the point $v t \mathsf s_1 $, the walk must make $s$ steps towards $\mathsf s_1$ and $t-s$ steps in other directions,  with $s \,\mathsf s_1\cdot\mathsf s_1 +(t-s) \mathsf s_1\cdot\mathsf s_2 = v t  \mathsf s_1\cdot\mathsf s_1$. Since $\mathsf s_1.\mathsf s_2=\frac{-1}{d+1}$, we obtain that 
\be
s=t\frac{1+vd}{d+1}.
\ee 
It is convenient to consider a biased random walk,  where we multiply the probability $p_1$, the probability to make the step $\mathsf s_1$,  by $\frac{e^a}{C}$ for some normalization constant $C=\overline{p_1}e^a+ 1-\overline{p_1}$. We have 
\be 
\frac{e^{a t\frac{1+vd}{d+1} }}{\left(\overline{p_1}e^a+ 1-\overline{p_1}\right)^t } \overline{\mathbf P(\mathsf X(t)=t v \mathsf s_1)} = \mathrm P\left( \widetilde X(t) = v t \mathsf s_1\right)
\ee 
where $ \widetilde X(t) $  is the biased random walk with transition probabilities 
\be 
\mathrm P(\widetilde X(t)=\widetilde X(t-1)+\mathsf s_i) = \frac{\overline{p_i}e^{a\mathds{1}_{i=1}}}{\overline{p_1}e^a+1-\overline{p_1}}.
\ee 
We chose $a$ so that $ \widetilde X(t) $ has drift $v$. Taking into account that $\sum_{i=1}^{d+1}\mathsf s_i =0$, we find 
\be 
v=\frac{e^a -1}{e^a+d}.
\ee 
Then, using the same notations as in \eqref{eq:localCLT}, as $t$ goes to infinity, we have 
\be  \mathrm P\left( \widetilde X(t) = v t \mathsf s_1\right)\sim \frac{1}{\sqrt{2\pi t \widehat{\det} \widetilde{D}}^d}
\ee 
where $\widetilde{ D}$ is the diffusion matrix of the walk $\widetilde X$. 
Finally, we have obtained that 
\be 
\boxed{\overline{\mathbf P(\mathsf X(t)=t v \mathsf s_1)} \sim\frac{e^{-t I(a)}}{\sqrt{2\pi t \widehat{\det} \widetilde{D}}^d}, \text{ where } I(a):=a\frac{e^a}{e^a+d} -\log\left( \frac{e^a+d}{d+1}\right).}
\ee

\subsection{Second moment}
Now we use the fact that $\overline{\mathbf P(\mathsf X(t)=\mathsf x)^2} = P(X_1(t) = X_2(t) =\mathsf x)$ where the couple of random walks $(X_1,X_2)$ has transition probabilities 
\be 
\mathrm P(X_1(t)=\mathsf x_1+\mathsf s_i, X_2(t)=\mathsf x_2+\mathsf s_j \Big\vert X_1(t-1)=\mathsf x_1, X_2(t-1)=\mathsf x_2 ) =\begin{cases}
\overline{p_i}\,\, \overline{p_j} &\mbox{ if }\mathsf x_1\neq \mathsf x_2, \\
\overline{p_i p_j} &\mbox{ if }\mathsf x_1 =  \mathsf x_2. 
\end{cases}
\ee 
In the notations above, we have used the fonts $X_1,X_2$ instead of $\mathsf X_1, \mathsf X_2$ used in previous sections, because the transition probabilities of $\mathsf X_1, \mathsf X_2$ depend on the environment, while here we have averaged over the environment in the probability distribution of the couple of random walks. 

As before, it will be convenient to use biased random walks. Following \cite{hass2025universal} (which is based on the special case treated in \cite{das2024multiplicative}), we consider the couple of random walks $(\widetilde{X_1}(t), \widetilde{X_2}(t))$
which are random walks starting from $0$ which evolve according to the following rules. If $\widetilde{X}_1(u)=\widetilde{ X}_2(u)$, then  $\widetilde{ X}_1(u+1)=\widetilde{ X}_1(u)+\mathsf s_i$   and $\widetilde{ X}_2(u+1)=\widetilde{ X}_2(u)+\mathsf s_j$ with probability 
\be 
\frac{\overline{p_i p_j} e^{a(\mathds{1}_{i=1} +\mathds{1}_{j=1}) }   }{\sum_{ 1\leq k,\ell\leq d+1}\overline{p_{k} p_{\ell}} e^{a(\mathds{1}_{k=1} +\mathds{1}_{\ell=1})  }}; 
\label{eq:transitionprobasame}
\ee 
while if $\widetilde{ X}_1(u)\neq \widetilde{ X}_2(u)$, then  $\widetilde{ X}_1(u+1)=\widetilde{ X}_1(u)+\mathsf s_i$   and $\widetilde{ X}_2(u+1)=\widetilde{ X}_2(u)+\mathsf s_j$ with probability 
\be 
\frac{\overline{p_i }\, \overline{p_j} e^{a(\mathds{1}_{i=1} +\mathds{1}_{j=1}) }   }{\sum_{ 1\leq k,\ell\leq d+1}\overline{p_{k} }\,\,\overline{p_{\ell}} e^{a(\mathds{1}_{k=1} +\mathds{1}_{\ell=1} )  }}.
\label{eq:transitionprobaapart}
\ee 
Now, we consider the asymptotics of the rescaled probability  
\be 
\left(\frac{e^{a t\frac{1+vd}{d+1} }}{\left(\overline{p_1}e^a+ 1-\overline{p_1}\right)^t }\right)^2 \overline{\mathbf P(\mathsf X(t)=t v \mathsf s_1)^2}, 
\label{eq:secondmomenttocompute}
\ee
but we need to take into account one subtlety. 
The denominator in the transition probabilities of the biased random walks in \eqref{eq:transitionprobaapart} is equal to the denominator in \eqref{eq:secondmomenttocompute}, i.e.,
\be 
\sum_{ 1\leq k,\ell\leq d+1}\overline{p_{k} }\,\,\overline{p_{\ell}} e^{a(\mathds{1}_{k=1} +\mathds{1}_{\ell=1} )  }  = \left(\overline{p_1}e^a+ 1-\overline{p_1}\right)^2.
\ee 
However, the denominator in the transition probabilities \eqref{eq:transitionprobasame} is not equal to 
the $\left(\overline{p_1}e^a+ 1-\overline{p_1}\right)^2$. Taking this into account, following \cite{hass2025universal, das2024multiplicative}, we find that 
\be 
\left(\frac{e^{a t\frac{1+vd}{d+1} }}{\left(\overline{p_1}e^a+ 1-\overline{p_1}\right)^t }\right)^2 \overline{\mathbf P(\mathsf X(t)=t v \mathsf s_1)^2} = \mathrm E\left[ \mathds{1}_{\widetilde X_1(t)=\widetilde X_2(t)=t v \mathsf s_1} \exp\left(g(a) \sum_{u=0}^{t-1}  \mathds{1}_{\widetilde{ X}_1(u)=\widetilde{ X}_2(u)} \right)  \right]
\label{eq:exponentiallocaltime}
\ee
where $\mathrm E$ denotes the expectation with respect to the law of the couple $(\widetilde{X_1}(t), \widetilde{X_2}(t))$ described above, and 
the function $g$ is given by 
\be 
g(a) = \log \frac{\overline{ \left( \sum_{i =1}^{d+1} p_ie^{a \mathds{1}_{i=1} } \right)^2} }{\left( \overline{  \sum_{i=1}^{d+1} p_ie^{a \mathds{1}_{i=1}} }  \right)^2} =\log \frac{\overline{(p_1(e^a-1)+1)^2}}{(\overline{p_1}(e^a-1)+1)^2} . 
\ee

Using $v=(e^a-1)/(e^a+d)$, this simplifies to
\be 
e^{g(a)} = 1 + \frac{d v^2}{(d+1)\alpha+1}
\label{eq:expressiong}
\ee

The conclusion of this section is that the second moment of $\mathcal Z_t$ is expressed as 
\be
\boxed{\overline{\mathcal Z_t^2} = \frac{1}{\mathrm P\left( \widetilde X(t) = v t \mathsf s_1\right)^2} \mathrm E\left[ \mathds{1}_{\widetilde X_1(t)=\widetilde X_2(t)=t v \mathsf s_1} \exp\left(g(a) \sum_{u=0}^{t-1}  \mathds{1}_{\widetilde{ X}_1(u)=\widetilde{ X}_2(u)} \right)  \right] }
\label{eq:exactsecondmoment}
\ee 
\subsection{Analysis of the local time}
As in the theorem of Erd\"{o}s-Taylor \cite{erdos1960some}, as $t$ goes to infinity, in dimension $d\geq 3$,  we will show that the local time $ \sum_{u=0}^{t-1} \mathds{1}_{\widetilde{ X}_1(u)=\widetilde{ X}_2(u)}$ converges to a geometric random variable, and we need to determine its parameter. By decomposing the trajectory of the random walk $\widetilde{ X}_1-\widetilde{ X}_2$ in excursions separated by periods where the walk stays at zero, we obtain the equality in distribution 
\be 
\sum_{u=0}^{\infty}  \mathds{1}_{\widetilde{ X}_1(u)=\widetilde{ X}_2(u)} = \sum_{i=1}^{1+Geom(\tilde Q)} (1+G_i )
\label{eq:geometricsum}
\ee 
where we have used the following notations. 
\begin{itemize} 
\item The number of excursions is distributed as $\mathrm{Geom}(\tilde Q)$; 
\item The parameter $\tilde Q$ is such that $1-\tilde Q$ is the probability that the walk $\mathsf Y:=\widetilde{ X}_1-\widetilde{ X}_2$ never returns to $0$, when starting from one of the points $\mathsf s_i-\mathsf s_j$ (neighbours of $0$); 
\item The variable $G_i$ denotes the length of the stay at $0$ between the $(i-1)$th excursion and the $i$-th excursion. The random variables $G_i$ are independent and distributed as $\mathrm{Geom}(\tilde q)$ 
where  
\be 
\tilde q = 
  \frac{\sum_{i=1 }^{d+1}\overline{p_i p_i} e^{2 a\mathds{1}_{i=1}  }   }{\sum_{1\leq  k,\ell \leq d+1}\overline{p_{k} p_{\ell}} e^{a(\mathds{1}_{k=1} +\mathds{1}_{\ell=1} ) }} = \frac{\frac{\alpha+1}{(d+1)((d+1)\alpha+1)}(e^{2a}+d) }{\frac{\alpha+1}{(d+1)((d+1)\alpha+1)}  (e^a-1)^2 + \frac{2}{d+1}(e^a-1) + 1 }.
\ee 
In other terms, $\tilde q$ is the probability for the walk  $\mathsf Y$ starting from $0$ to  stay at zero after one step.  For example, when $d=3$, we find, 
\be 
\tilde q = \frac{\frac{\alpha+1}{4(4\alpha+1)} (e^{2a}+3)}{\frac{\alpha+1}{4(4\alpha+1)}(e^a-1)^2  +\frac{1}{2}(e^a-1) +1}.
\ee 
\end{itemize} 
We conclude from \eqref{eq:geometricsum} that 
\be 
\sum_{u=0}^{\infty}  \mathds{1}_{\widetilde{ X}_1(u)=\widetilde{ X}_2(u)} \sim 1+\mathrm{Geom}(R), \text{ where } R = 1-(1-\tilde q)(1- \tilde Q).
\ee
However, in \eqref{eq:exactsecondmoment}, we need to take into account the indicator function. 
It is convenient to approximate the expectation in \eqref{eq:exactsecondmoment} as the conditional expectation
\be 
\mathrm E\left[\exp\left(g(a) \sum_{u=0}^{t-1}  \mathds{1}_{Y(u)=0} \right)\Big\vert Y(t)=0  \right] \mathrm P(\widetilde X_1(t)=\widetilde X_2(t)=t v \mathsf s_1).
\ee 
so that as $t$ goes to infinity, 
\be \boxed{
\overline{\mathcal Z_t^2}
   \sim \mathrm E\left[\exp\left(g(a) \sum_{u=0}^{t-1}  \mathds{1}_{Y(u)=0} \right)\Big\vert Y(t)=0  \right] \frac{\mathrm P(\widetilde X_1(t)=\widetilde X_2(t)=t v \mathsf s_1)}{\mathrm P(\widetilde X(t)=t v \mathsf s_1)^2}.}
   \label{eq:secondmomentestimate}
   \ee
The last ratio converges to a constant as $t$ goes to infinity, which we will compute below. The conditional expectation, however, exhibits a transition. 

\subsubsection{Analysis of the conditioned local time }
The distribution of the local time for the random walk $\mathsf Y$ conditioned to return to $0$ at time $t$ has the distribution of the sum of two independent copies of the unconditioned local time (that is a geometric random variable with parameter $R = 1-(1-\tilde q)(1- \tilde Q)$ as we have seen above).

Indeed, roughly speaking,  there is one contribution to the local time coming from the portion of the walk starting from $0$ before it escapes at infinity, and another independent contribution from the portion of the walk which arrives at zero at time $t$. 
Hence, 
the rescaled second moment remains bounded as $t$ goes to infinity as long as $v$ is chosen such that 
\be 
 R= 1- (1-\tilde q)(1-\tilde Q) <e^{- g(a)}.
\ee 

We now need to compute $\tilde Q$. We use the fact that this parameter is also related to the number of excursions in the modified walk $\overline{ \mathsf Y } $ (that is, as in Appendix \ref{sec:thermalaverage}, the modification of the walk $\mathsf Y$ that has the same transition probabilities at zero and away from zero). More precisely, by applying to the walk $\overline {\mathsf Y}$ the same decomposition in excursions as we have done just above for the walk $\mathsf Y$, the total local time of $\overline{ \mathsf Y }$  has the same distribution as  $1+\mathrm{Geom}(1-(1-\bar q)(1- \tilde Q)) $ where $\tilde q$ has been replaced by $\bar q$ with 
\be 
\bar q = 
  \frac{\sum_{i }\overline{p_i} \,\,\overline{p_i} e^{2 a\mathds{1}_{i=1}  }   }{\sum_{ k,\ell}\overline{p_{k}}\,\, \overline{ p_{\ell}} e^{a(\mathds{1}_{k=1} +\mathds{1}_{\ell=1} ) }} = \frac{ (e^{2a}+ d)}{(e^a+d)^2}.
\ee
For example,  when $d=3$, we find 
\be  
\bar q
= \frac{3 + e^{2 a}}{(3+ e^a)^2}.
\ee  
We will compute $\tilde Q$ through the expected total local time of the walk $\overline{ \mathsf Y } $, that we denote by 
\be L(a) = \mathrm E\left[ \sum_{t=0}^{\infty} \mathds{1}_{\overline{\mathsf Y}(t)=0}\right]. 
\ee 
Indeed, since the total local time is geometrically distributed with parameter $1-(1-\bar q)(1-\tilde Q)$, 
we have 
\be 
L(a)= \frac{1}{(1-\bar q)(1-\tilde Q)}.  
\ee 
This allows to write the geometric parameter $R(a)$ in terms of the expected local time $L(a)$: 
\begin{align}
R(a) &= 1- (1-\tilde q)(1-\tilde Q)\\
&= 1- \frac{1-\tilde q}{1-\bar q} \frac{1}{L(a)}\\ 
&= 1- \frac{1}{L(a)} \frac{(d+1)\alpha}{(d+1)\alpha+1+d v^2}.
\end{align} 
This implies that  the second moment remains bounded as long as 
\be   
 1- \frac{1}{L(a)} \frac{(d+1)\alpha}{(d+1)\alpha+1+d v^2} <e^{-g(a)}  . \ee 
Using \eqref{eq:expressiong},  the condition for a finite second moment simplifies and becomes 
\be  \boxed{L(a)< \frac{(d+1) \alpha}{d v^2}}
\label{eq:conditionsecondmoment}
  \ee
It remains to compute the expected local time $L=L(a)$ as a function of $a$. In dimension 3, by similar arguments as in \eqref{eq:integralreturnproba}, we have 
\begin{equation}L(a) =
    \frac{1}{(2\I\pi)^3}  \oint\frac{\mathrm dz_1}{z_1} \oint \frac{\mathrm dz_2}{z_2}\oint \frac{\mathrm dz_3}{z_3}  \frac{1}{1-f_a(z_1,z_2,z_3)}, 
    \label{La}
\end{equation}
where 
\begin{equation}  
f_a(z_1,z_2,z_3) = \frac{1}{(e^a+3)^2} \left( e^{a} + z_1 +z_2+z_3\right) \left( e^{a} + \frac{1}{z_1 }+ \frac{1}{z_2 }+ \frac{1}{z_3 }\right).
\end{equation}   

\subsubsection{Numerics}
Thus, the rescaled second moment \eqref{eq:rescaledsecondmoment} remains finite as $t$ goes to infinity as long as $v < v_2$ where $v_2$ is some critical value (which is such that $v_2\leq v_c$). Computing numerically the function $L(a)$, and recalling that $e^a = (1+ 3 v)/(1-v)$, we find that the critical value is 
\be 
\boxed{v_2= 0.639415}.
\ee 
Since we consider vectors $\mathsf u(k) = v \mathsf s_1 + \mathsf d$ where $\mathsf u(k) = (\frac{k-3}{k}, \frac{1}{k},\frac{1}{k},\frac{1}{k}) $, the velocity $v$ is related to $k$ by  
 \be 
v = 1 - \frac{4}{k}.
\ee 
This leads to a critical value of $k$ being $k_2=4/(1-v_2)\approx 11.09$.

\begin{remark}
      Small $\alpha$ behavior: Recalling the numerical value $L(0)=1.79288$ obtained in \eqref{eq:L(0)}, we get $v_2 \simeq \sqrt{4\alpha/(3 L(0))} \approx 0.86237 \sqrt{\alpha}$ at small $\alpha$.
\end{remark}

\subsection{Limit in the weak disorder phase}

In the weak disorder phase we expect that the distribution of the random variable 
$\mathbf P(\mathsf X(t)= v t \mathsf s_1)$ converges to a limit for $t \to +\infty$. 
From \cite{junk2024tail} (see also 
\cite{lacoin2025localization,junk2024strong,junk2025coincidence} ) 
we expect that this limit distribution has a power law tail 
with exponent $\mu(v) > \mu(v_c) = 1 + 2/d$, i.e.,$\mu(v_c)=5/3$ in $d=3$. 
The second moment thus reaches a finite limit at large $t$ for $\mu=\mu(v)>2$.  
We now estimate this limit.

Recall our approximation \eqref{eq:secondmomentestimate}, and the fact that the conditioned local time of $\mathsf Y$ is distributed as $1+ \mathrm{Geom}(R)+ \widetilde{\mathrm{Geom}}(R)$, where the two Geometric variables are independent. This implies that 
\be 
\overline{\mathcal Z_t^2}
   \sim  e^{g(a)}\left( \mathrm{E}\left[e^{g(a) \mathrm{Geom}(R)} \right] \right)^2 \frac{\mathrm P(\widetilde X_1(t)=\widetilde X_2(t)=t v \mathsf s_1)}{\mathrm P(\widetilde X(t)=t v \mathsf s_1)^2}, 
   \label{eq:estimatesecondmomentgeometric}
   \ee
where we recall that 
\be R=R(a):=1- \frac{1}{L(a)} \frac{(d+1)\alpha}{(d+1)\alpha+1+d v^2}.
\label{eq:expressionR}
\ee
The ratio of probabilities is controlled by the local central limit theorem satisfied by the couple of biased random walks $(\widetilde X_1(t), \widetilde X_2(t))$. We claim that its stationary measure\footnote{it is unique, up to a multiplicative constant, see \cite[Theorem 4.10]{drillick2025random}} is equal to 
\begin{equation}  \widetilde\nu(\mathsf x_1,\mathsf x_2) = 
\begin{cases}
1 & \mbox{ if }\mathsf x_1\neq \mathsf x_2  \\ 
\frac{\alpha(d+1)+1}{\alpha(d+1)} e^{g(a)} & \mbox{ if }\mathsf x_1=\mathsf x_2.
\end{cases} 
\label{eq:stationarymeasuretilted}
   \end{equation}
 This comes from the fact that the stationary measure for the couple of unbiased walks $(X_1,X_2)$ is the average over the noise of the stationary measure for random walks in random environment, which is given by the product $\gamma_{\mathsf x_1}\gamma_{\mathsf x_2}$ where $(\gamma_{\mathsf x})_{\mathsf x\in \mathbb T_d}$ is an i.i.d. family of $\mathrm{Gamma}(\alpha(d+1))$ random variables. Hence the measure 
 \be 
\nu(\mathsf x_1, \mathsf x_2) = \overline{\gamma_{\mathsf x_1}\gamma_{\mathsf x_2}} = \begin{cases}1 & \mbox{ if }\mathsf x_1\neq \mathsf x_2  \\ 
\frac{\alpha(d+1)+1}{\alpha(d+1)} & \mbox{ if }\mathsf x_1=\mathsf x_2.
\end{cases}
 \ee 
is stationary for the Markov process describing the unbiased walks $(X_1(t),X_2(t))$.  Then, the factor $e^{g(a)}$ in \eqref{eq:stationarymeasuretilted} comes from the tilt. To see precisely how this factor arises, recall that the stationary measure of the couple of biased random walks $(\widetilde X_1, \widetilde X_2)$ is, by definition, the solution (up to multiplicative constants) of the relation
 \be 
\sum_{1\leq i,j \leq d+1} \frac{ e^{a \mathds{1}_{i=1} + a \mathds{1}_{j=1} }}{(\overline{p_1}e^a+1-\overline{p_1})^2 } \overline{p^{i}_{\mathsf x-\mathsf s_i} p^{j}_{\mathsf y-\mathsf s_j}} e^{-\mathds{1}_{\mathsf x-\mathsf s_i=\mathsf y-\mathsf s_j}g(a)}  \widetilde\nu(\mathsf x-\mathsf s_i,\mathsf y-\mathsf s_j ) = \widetilde\nu(\mathsf x, \mathsf y), 
\label{eq:stationaryrelation}
 \ee 
 where the $p_{\mathsf x}=(p^1_{\mathsf x}, \dots, p^{d+1}_{\mathsf x})$ are i.i.d. Dirichlet vectors indexed by positions $\mathsf x\in \mathbb T_d$. 
 This suggests to let $\widetilde\nu(\mathsf x, \mathsf y)=e^{\mathds{1}_{\mathsf x=\mathsf y}g(a)}\nu(\mathsf x, \mathsf y)$ so that \eqref{eq:stationaryrelation} reduces to 
 \be 
\sum_{1\leq i,j \leq d+1} e^{a \mathds{1}_{i=1} + a \mathds{1}_{j=1} }  \overline{p^{i}_{\mathsf x-\mathsf s_i} p^{j}_{\mathsf y-\mathsf s_j}} \nu(\mathsf x-\mathsf s_i,\mathsf y-\mathsf s_j ) = \nu(\mathsf x, \mathsf y)\sum_{1\leq i,j \leq d+1} e^{a \mathds{1}_{i=1} + a \mathds{1}_{j=1} }  \overline{p^{i}_{\mathsf x} p^{j}_{\mathsf y}}. 
\ee 
 This last equation holds thanks to  some skew-detailed-balance identity satisfied by the measure $\nu$:  for all $1\leq i,j\leq d+1$, and for all $\mathsf x, \mathsf y\in \mathbb T_d$, 
 \be 
\overline{p^{i}_{\mathsf x-\mathsf s_i} p^{j}_{\mathsf y-\mathsf s_j}} \nu(\mathsf x-\mathsf s_i,\mathsf y-\mathsf s_j ) = \overline{p^{i}_{\mathsf x} p^{j}_{\mathsf y}}  \nu(\mathsf x, \mathsf y). 
\label{eq:skewreversibility}
\ee 

 We deduce from the form of the stationary measure \eqref{eq:stationarymeasuretilted} that 
   \be 
\lim_{t\to\infty} \frac{\mathrm P(\widetilde X_1(t)=\widetilde X_2(t)=t v \mathsf s_1)}{\mathrm P(\widetilde X(t)=t v \mathsf s_1)^2} = \widetilde\nu(t v \mathsf s_1, t v \mathsf s_1) = \frac{\alpha(d+1)+1}{\alpha(d+1)}e^{g(a)}.
\label{eq:limitratioproba}
\ee

Thus, using the estimation \eqref{eq:estimatesecondmomentgeometric}, the limit \eqref{eq:limitratioproba}, and the explicit expressions for $R(a)$ and $g(a)$ given in \eqref{eq:expressionR} and in \eqref{eq:expressiong}, we find that 
\begin{align}
m_2(v):= \lim_{t\to\infty} \frac{\overline{\mathbf P(\mathsf X(t)=t v \mathsf s_1)^2}}{\left(\overline{\mathbf P(\mathsf X(t)=t v \mathsf s_1)}\right)^2}  
&=  \left(  \frac{1-R(a)}{e^{-g(a)}-R(a)} \right)^2 \frac{\alpha(d+1)+1}{\alpha(d+1)} \\
&= \left(\frac{1}{1-\frac{dv^2 L(a)}{(d+1)\alpha}}\right)^2 \frac{\alpha(d+1)+1}{\alpha(d+1)}.
\end{align} 
In the case $d=3$, we get 
\be 
\boxed{ m_2(v) = 
\frac{
4\alpha(4\alpha+1)
}{
\left(4\alpha-3v^2 L(a)\right)^2
} = \frac{ 1 + \frac{1}{4 \alpha}}{(1- \frac{3v^2}{4 \alpha}  L(a))^2 }}
\ee 
where $L(a)$ is given in \eqref{La}. Using $e^a = (1+ 3 v)/(1-v)$ and $v=1-4/k$, we find the following table of numerical values, when $\alpha=1$:

\smallskip
\begin{center} 
\begin{tabular}{|c|c|c|c|c|c|c|c|c|c|c|c|c|}
\hline
    $k$  &  4&5&6&7&8&9&10&11 & 11.09...\\
    \hline 
 $m_2(\mathsf u(k))$    & 1.25& 1.4051& 1.82882& 2.67118& 4.49578& 9.53944& 34.2218&4639.45 &$+\infty$\\
\hline
\end{tabular}
\end{center}

\begin{remark}
The same transition can also be understood from a resolvent point of view. Consider the difference walk \(Y(u)=\widetilde X_1(u)-\widetilde X_2(u)\), and let \(A\) denote the homogeneous transition kernel away from the origin. The effect of the contact reward \(e^{g(a)}\), together with the modified transition probabilities when \(Y=0\), is a rank-one perturbation of this homogeneous kernel, so that the transfer operator can be written as
\[
M=A+|B\rangle\langle 0|.
\]
The resolvent is then obtained from the Sherman--Morrison formula,
\[
(zI-M)^{-1}
=
(zI-A)^{-1}
+
\frac{(zI-A)^{-1}|B\rangle\langle 0|(zI-A)^{-1}}
{1-\langle 0|(zI-A)^{-1}|B\rangle}.
\]
Thus the loss of boundedness of the second moment corresponds to the appearance of a pole of the resolvent at \(z=1\), or equivalently to the formation of a bound state localized near the origin for the difference walk. The criticality condition is therefore
\[
1=\langle 0|(I-A)^{-1}|B\rangle,
\]
which, after Fourier transform, is equivalent to the local-time criterion
\[
L(a)=\frac{4\alpha}{3v^2}
\]
in dimension \(d=3\), with \(e^a=(1+3v)/(1-v)\). This gives the same divergence threshold as the geometric-local-time computation.
\end{remark}

\section{Saul-Kardar-Read's model in higher dimension}
\label{sec:SKR}
In this section, we consider a discrete version of the unitary evolution 
\be 
\I\partial_t\psi(x,t) = (-\Delta + \xi(x,t)) \psi(x,t), \;\; t\in \R_+, x\in \R^d.
\label{eq:directedwaves}
\ee 
In the discrete model, the wavefunction takes values in $\mathbb C^{d+1}$ and is defined on $\Z^{d+1}$ which we may identify with $\mathbb T_d\times \Z$ (we recall that $\mathbb T_d$ is the hyperplane orthogonal to $\mathsf d$). Hence, we consider a wavefunction 
\be 
\Psi(\mathsf x,t) = \begin{pmatrix}
    \Psi_1(\mathsf x,t) \\ \vdots \\ \Psi_{d+1}(\mathsf x,t)
\end{pmatrix} \in \mathbb C^{d+1}, \;\;\;\mathsf x\in \mathbb T_d
\label{eq:coordinatesPsi}
\ee 
To define the evolution, it is more convenient to work with coordinates $\mathsf n\in \Z_+^{d+1}$ rather than coordinates $(\mathsf x,t)\in \mathbb T_d\times \Z_+$. We will use the notation $\Psi(\mathsf n) = \Psi(\mathsf x,t)$ when $\mathsf x=\mathsf n-t\mathsf d$, as above. The evolution of the wave function is given by  
\be 
\begin{pmatrix}
    \Psi_1(\mathsf n+\mathsf e_1) \\ \vdots \\ \Psi_{d+1}(\mathsf n+\mathsf e_{d+1})
\end{pmatrix}= U(\mathsf n)  \begin{pmatrix}
    \Psi_1(\mathsf n) \\ \vdots \\ \Psi_{d+1}(\mathsf n )
\end{pmatrix}
\ee
where $U(\mathsf n)$ are independent Haar distributed $(d+1)\times(d+1)$ unitary matrices. For a Haar distributed random matrix $U$, any column is distributed as a uniform vector on the complex sphere. This implies that, say for the first column, the moduli squared of the matrix components are distributed as a Dirichlet vector, i.e.,
\be 
\left( \vert U_{11}\vert^2, \dots, \vert U_{(d+1),1 } \vert^2\right) \sim \mathrm{Dir}(1,\dots, 1).
\ee 
Using the translation invariance of the Haar distribution, we deduce that at each step, the update is such that 
\be 
\vert \Psi_k(\mathsf n+\mathsf e_k) \vert^2 = p^k_{\mathsf n}  \sum_{i=1}^{d+1} \vert \Psi_i(\mathsf  n) \vert^2
 \ee 
where the probabilities $p^k$ are Dirichlet distributed exactly as in the main text, with $\alpha= 1$. Intuitively, $\vert \Psi_i(\mathsf n) \vert^2$ represents the probability for a quantum particle with wavefunction $\Psi$ to arrive at the site $\mathsf n$ coming from direction $\mathsf e_i$. This implies that if the wavefunction is initially localized with $\Vert \Psi(0) \Vert=1$, we have 
\be 
\Vert \Psi(\mathsf n) \Vert^2 = P(\mathsf n).
\ee 
\begin{remark}
The random evolution of $\Psi(\mathsf x,t)$ may be seen as a Markov process. Its stationary measure is such that   $\Psi(\mathsf x, t)$ are independent standard Gaussian variables in $\mathbb C^{d+1}$. Indeed, the squared norm of the stationary wave function should be distributed as the stationary measure of the Dirichlet RWRE. Unitary  invariance then implies that the distribution is exactly Gaussian. 
\end{remark}
\begin{remark}
When $\alpha_1, \dots, \alpha_{d+1}$ are positive integers, one can likewise relate the $\mathrm{Dir}(\boldsymbol\alpha)$-RWRE to the evolution of a wave function $\Psi$. The model is the same as above except that  now, the wavefunction $\Psi$ is  $\mathbb C^D$-valued, where $D=\alpha_1+\dots+\alpha_{d+1}$. One has to assume in \eqref{eq:coordinatesPsi} that $\Psi_i \in \mathbb C^{\alpha_i}$ and the matrices $U$ are now Haar distributed matrices in $U(D)$. 
\end{remark}
\begin{remark}
The connection between RWRE and the KPZ equation suggests the following open problem:   consider a solution to  the directed waves equation \eqref{eq:directedwaves} -- regularized in a certain way, to be determined.  In the moderate deviations scaling, i.e.,as in \cite{ledoussal2017diffusion}, can one show that the modulus squared of $\psi$ converges to the KPZ equation.
\end{remark}

\end{widetext}

\end{document}